\documentclass[]{aastex63}

\accepted{June 23, 2023}
\submitjournal{AJ}

\shorttitle{M-Dwarf UV Spectra with UVIT}
\shortauthors{Ranjan et al.}
\graphicspath{{./}{figures/}}

\begin{document}

\title{UV Spectral Characterization of Low-Mass Stars With AstroSat UVIT for Exoplanet Applications: The Case Study of HIP 23309}

\correspondingauthor{Sukrit Ranjan}
\email{sukrit@arizona.edu}

\author[0000-0002-5147-9053]{Sukrit Ranjan}
\altaffiliation{Center for Interdisciplinary Exploration and Research in Astrophysics \& Department of Physics and Astronomy, Northwestern University, Evanston, 60201, USA}
\altaffiliation{Indian Institute of Astrophysics, Bengaluru, 560034, India}
\affiliation{University of Arizona, Lunar and Planetary Laboratory/Department of Planetary Sciences, Tucson, 85721, USA}

\author[0000-0002-4638-1035]{Prasanta K. Nayak}
\affiliation{Institute of Astrophysics, Pontificia Universidad Católica de Chile, Av. Vicuña MacKenna 4860, 7820436, Santiago, Chile}
\affiliation{Department of Astronomy and Astrophysics, Tata Institute for Fundamental Research, Mumbai, 400005, India}

\author[0000-0001-8270-2476]{J. Sebastian Pineda}
\affiliation{University of Colorado-Boulder, Boulder, 80309, USA}

\author[0000-0002-0554-1151]{Mayank Narang}
\affiliation{Academia Sinica Institute of Astronomy \& Astrophysics, 11F of Astro-Math Bldg., No.1, Sec. 4, Roosevelt Rd., Taipei 10617, Taiwan, R.O.C.}
\affiliation{Department of Astronomy and Astrophysics, Tata Institute for Fundamental Research, Mumbai, 400005, India}



\begin{abstract}

Characterizing rocky exoplanet atmospheres is a key goal of exoplanet science, but interpreting such observations will require understanding the stellar UV irradiation incident on the planet from its host star. Stellar UV mediates atmospheric escape, photochemistry, and planetary habitability, and observations of rocky exoplanets can only be understood in the context of the UV SED of their host stars. Particularly important are SEDs from observationally favorable but poorly understood low-mass M-dwarf stars, which are the only plausible targets for rocky planet atmospheric characterization for the next 1-2 decades. In this work, we explore the utility of AstroSat UVIT for the characterization of the UV SEDs of low-mass stars. We present observations of the nearby M0 star HIP 23309 in the FUV and NUV gratings of UVIT. Our FUV spectra are consistent with contemporaneous HST data and our NUV spectra are stable between orbits, suggesting UVIT is a viable tool for the characterization of the SEDs of low-mass stars. We apply our measured spectra to simulations of photochemistry and habitability for a hypothetical rocky planet orbiting HIP 23309 and elucidate the utility and limitations of UVIT in deriving UV SEDs of M-dwarf exoplanet hosts. Our work validates UVIT as a tool to complement HST in the characterization of exoplanet host stars and carries implications for its successor missions like INSIST.

\end{abstract}



\section{Introduction}
Temperate, rocky planets orbiting other stars are abundant, and upcoming facilities will have the ability to characterize their atmospheres \citep{Dressing2015, Bryson2021, Rodler2014, Morley2017, Lustig-Yaeger2019}. Characterization of rocky exoplanet atmospheres will probe their surface processes including potentially constraining the presence or absence of life \citep{Schwieterman2018}, and such constraints are high priorities for the astronomical community \citep{NAP25187, Decadal2020}. However, interpreting measurements of rocky exo-atmospheres requires understanding the UV spectral energy distribution (SED) of their host stars. Stellar UV is a first-order control on the surface-atmosphere system of rocky planets, and mediates atmospheric escape, photochemistry, and habitability, meaning that observations of rocky exo-atmospheres can only be understood in the context of the SED of their host star \citep{Teal2022}. This observation has motivated observational and theoretical work aimed at constraining the high-energy emission of exoplanet host stars (e.g., \citealt{France2013, Viswanath2020}). 

Characterization of the UV emission of low-mass ($0.08-0.6~M_\sun$) M-dwarf stars is of particular importance to the quest to characterize rocky exo-atmospheres. Most stars are M-dwarfs, and planets orbiting M-dwarfs are uniquely accessible to characterization \citep{Scalo2007, Shields2016rev}. Consequently, M-dwarf planets are the \emph{only} plausible targets for in-depth atmospheric characterization and biosignature search with near-term facilities such as JWST and the ELTs \citep{Cowan2015, Morley2017, Batalha2018, Tremblay2020AJ....159..117T}. However, M-dwarf UV emission is very different from the UV emission of higher-mass stars. These differences are predicted to impact atmospheric escape, atmospheric photochemistry, and habitability relative to Sun-like stars \citep{Segura2005, RamirezKaltenegger2014, Luger2015, Rugheimer2015mdwarf, Dong2017, Rimmer2018, Wordsworth2018, Harman2018, Tilley2019, Schwieterman2022}, motivating observational efforts to characterize the SEDs of M-dwarf stars both directly (e.g., \citealt{France2013, France2016, Pineda2021}) and by using measurements to calibrate models (e.g., \citealt{Peacock2019a, Duvvuri2021, Diamond-Lowe2021}). Such observations must be done from space due to telluric absorption, and to-date are primarily executed on the Hubble Space Telescope (HST) (e.g., \citealt{France2016}). However, the energetic emission of M-dwarfs remains underconstrained, in part because of the need to use extremely scarce HST time, and in part, because the considerable diversity of M-dwarfs necessitates a large number of observations to constrain their emission as a function of parameters such as mass and age \citep{Loyd2021ApJ...907...91L, Pineda2021}. Efforts have been made to predict M-dwarf UV SEDs from more tractably-measured emission lines \citep{Youngblood2017, Melbourne2020}, but the use of such approximations may lead to significant inaccuracies in models of rocky exoplanet atmospheres. It remains preferred to characterize the UV SEDs of planet-hosting M-dwarf stars to accurately model the atmospheres of their orbiting planets \citep{Teal2022}.

The UltraViolet Imaging Telescope (UVIT) on the {Indian Space Research Organization (ISRO) Astronomy Satellite (AstroSat) mission is in principle well-suited to the characterization of the UV SEDs of M-dwarf stars. UVIT is composed of two 38-cm telescopes with a field of view of 28$'$ in diameter. One telescope is dedicated to observations in the FUV region (130-180 nm), while the other one is for NUV (200-300 nm) and visible regions (320-550 nm). The UVIT operates in photon counting mode and observes simultaneously with both the telescopes with a spatial resolution (FWHM) $<$1.$''$5 \citep{tandon2020}, better than the resolution of GALEX ($\sim5''$). Both FUV and NUV channels incorporate multiple narrow to broad-band filters. UVIT also has a facility for low-resolution slit-less spectroscopy in both NUV and FUV bands: one grating is mounted in the NUV wheel (now inoperative) and two gratings (with orthogonal dispersions axis) in the FUV wheel. The maximum efficiency is achieved in the $-$2 order of the FUV gratings with a spectral resolution of 15 \AA~and $-$1 order of the NUV grating with a spectral resolution of 33 \AA~\citep{tandon2020, Dewangan2021}. The NUV channel is now inoperative and only archival studies are possible; the FUV channel remains operative.

In this paper, we explore the utility of UVIT to constrain the SEDs of M-dwarf stars using HIP 23309 as a case-study. HIP 23309 is a young ($\sim24$ Myr), rapidly rotating, early-type M-dwarf star (\citealt{Pineda2021}; Table~\ref{tab:hip23309_params}). The age estimate for HIP 23309 is based on its membership in the $\beta$ Pictoris moving group \citep{Messina2010, Pineda2021}, which is $24\pm3$ Myr old \citep{Bell2015}. We secure spectrally resolved observations of HIP 23309 in the UVIT FUV ($130-180$ nm) and NUV ($200-300$ nm) gratings. Standard UVIT calibration pipelines are able to extract resolved spectra of HIP 23309 with no modification \citep{Dewangan2021}. We compare our FUV data to contemporaneous HST measurements and find agreement at levels that we consider satisfactory for use in exoplanet photochemistry studies. We perform atmospheric and surficial photochemical modeling to elucidate the utilities and limitations of such measurements for studying the atmospheres of rocky planets and assessing their habitability. We close by commenting on the potential of UVIT for studies of M-dwarfs and their exoplanets, and recommendations for future facilities like INSIST \citep{Subramaniam2022}.

 \begin{deluxetable}{lcr}
\tablecaption{HIP 23309 Stellar Parameters\label{tab:hip23309_params}}
\tablewidth{0pt}
\tablehead{
\colhead{Parameter} & \colhead{Value} & \colhead{Reference}
 }
\startdata
Spectral Type & M0& \citet{Torres2006, Pineda2021} \\
Effective Temperature & $3886^{+28}_{-27}$ K & \citet{Pineda2021} \\
Mass & $0.785^{+0.009}_{-0.010}$ $M_\Sun$ & \citet{Pineda2021} \\
Radius & $0.932\pm0.014$ $R_\Sun$ & \citet{Pineda2021} \\
Luminosity & $6.82^{+0.12}_{-1.1}\times10^{32}$ erg s$^{-1}$ & \citet{Pineda2021} \\
Rotation Period & $8.6\pm0.07$ days & \citet{Messina2010}\\
\enddata
\end{deluxetable}

\section{Observations and Data Reduction}
We observed HIP 23309 with the FUV-Grating1 (FUV-G1) and NUV-Grating of UVIT as a part of the AO-1 proposal (PI: Lalitha Sairam) on 24-11-2017 with an exposure time of $\sim$12 ksec in each channel. 
The observations were completed in multiple orbits. We applied corrections for spacecraft drift, flat-field, and distortion,  using the software CCDLAB \citep{postma2017} and created images for each orbit. Then, the orbit-wise images were co-aligned and combined to generate final science-ready images in each grating. Astrometry on the images was also performed using CCDLAB by comparing the Gaia-DR2 source catalogue. The zoomed-in FUV and NUV grating images are shown in \autoref{image}, with boxes indicating the different spectral orders. We extracted the source spectrum of $-$2 order from FUV-G1 image and $-1$ order from the NUV image with a cross-dispersion
width of 50 pixels (1 pixel $\sim$ $0.''41$). Background regions are selected exactly in the same pixel range above the source spectrum along the dispersion axis for both gratings. We used the UVITTools package written in the Julia language by \citealt{Dewangan2021} to extract spectral files for a given grating order and cross-dispersion width. Wavelength and flux calibration to the spectra are done using standard UVIT source calibrators: HZ4 and NGC40. We re-ran the pipeline on the raw HZ4 and NGC40 data and recovered results consistent with IUE measurements and \citet{Dewangan2021}, confirming that we were not using the pipeline erroneously.
We show the FUV and NUV spectra of HIP~23309 in \autoref{lines_uvit}.

\begin{figure}[ht!]
\plottwo{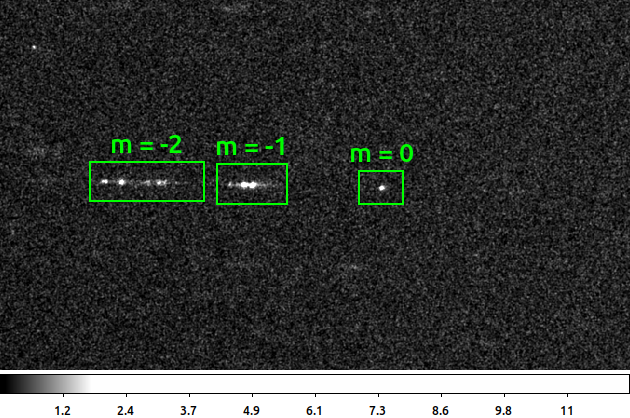}{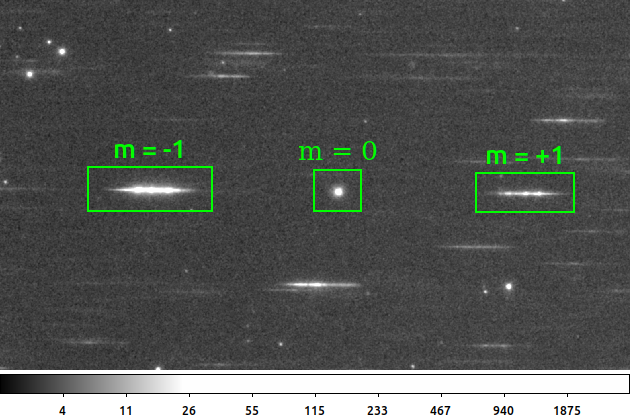}
\caption{Images of HIP 23309 in FUV-Grating1 (left) and NUV-Grating (right) are shown here. Different grating orders are marked in green. \label{image}}
\end{figure}

\section{Measured Spectra of HIP 23309}
\subsection{UVIT Measurements of HIP 23309}

In the NUV spectrum (right panel of Figure~\ref{lines_uvit}), we can clearly identify three peaks: one at 2800 \AA~due to strong Mg\textsc{ii} \textit{h} and \textit{k} lines, and the other two peaks at 2400 \AA~and 2600 \AA~due to strong Fe\textsc{ii} lines. The strong Mg\textsc{ii} \textit{h} and \textit{k} lines emissions indicate that the star is chromospherically active \citep{1979Natur.280..661B,1985MNRAS.217...41D}. 
In the FUV spectrum (left panel of \autoref{lines_uvit}), comparative poor signal-to-noise (SNR) makes it difficult to identify emission lines. We can clearly identify the strong C\textsc{iv} line at 1560 \AA~despite low SNR, however, the identification of  He\textsc{ii} and O\textsc{iii} are less reliable for further analysis.

\begin{figure}[ht!]
\plottwo{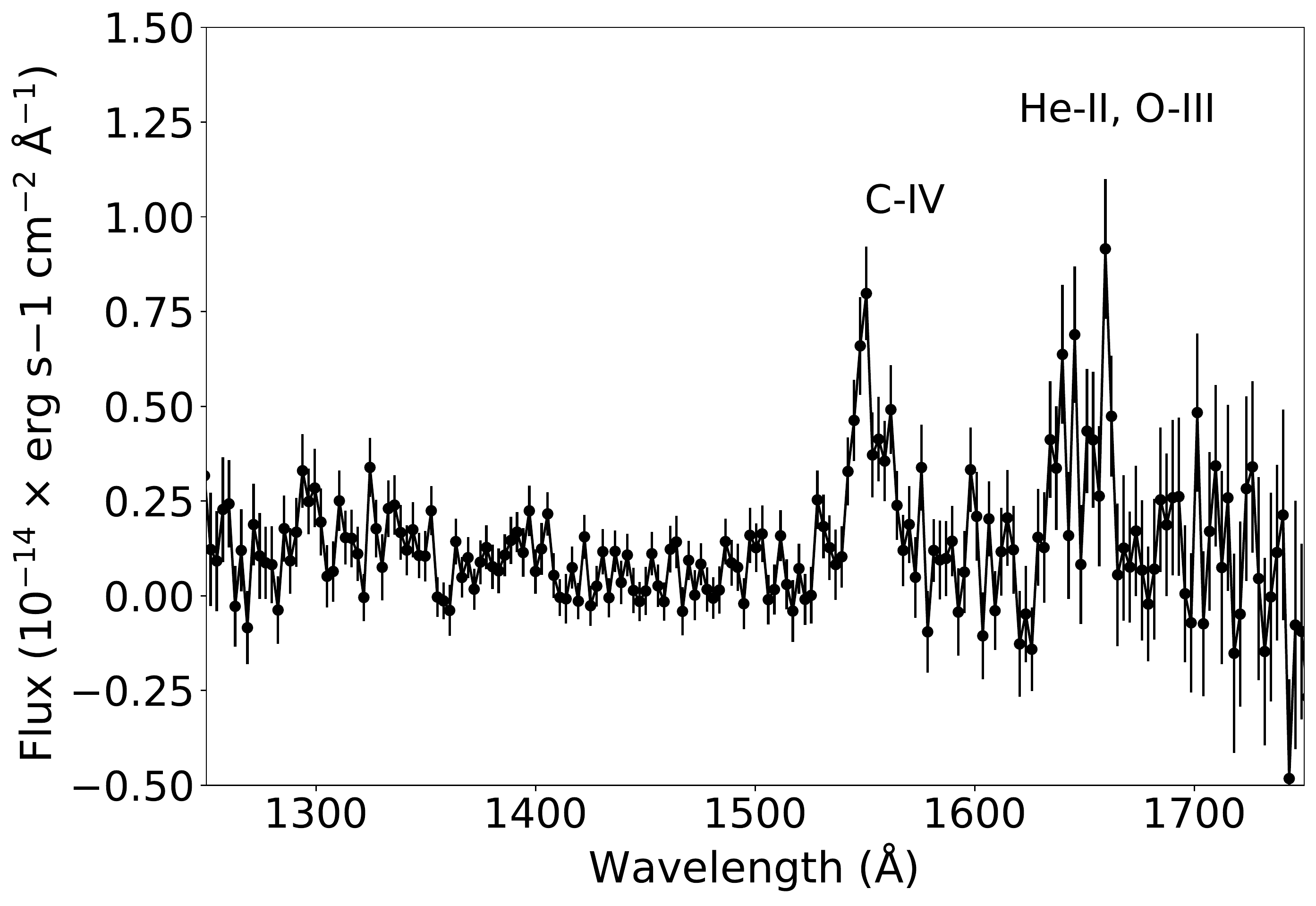}{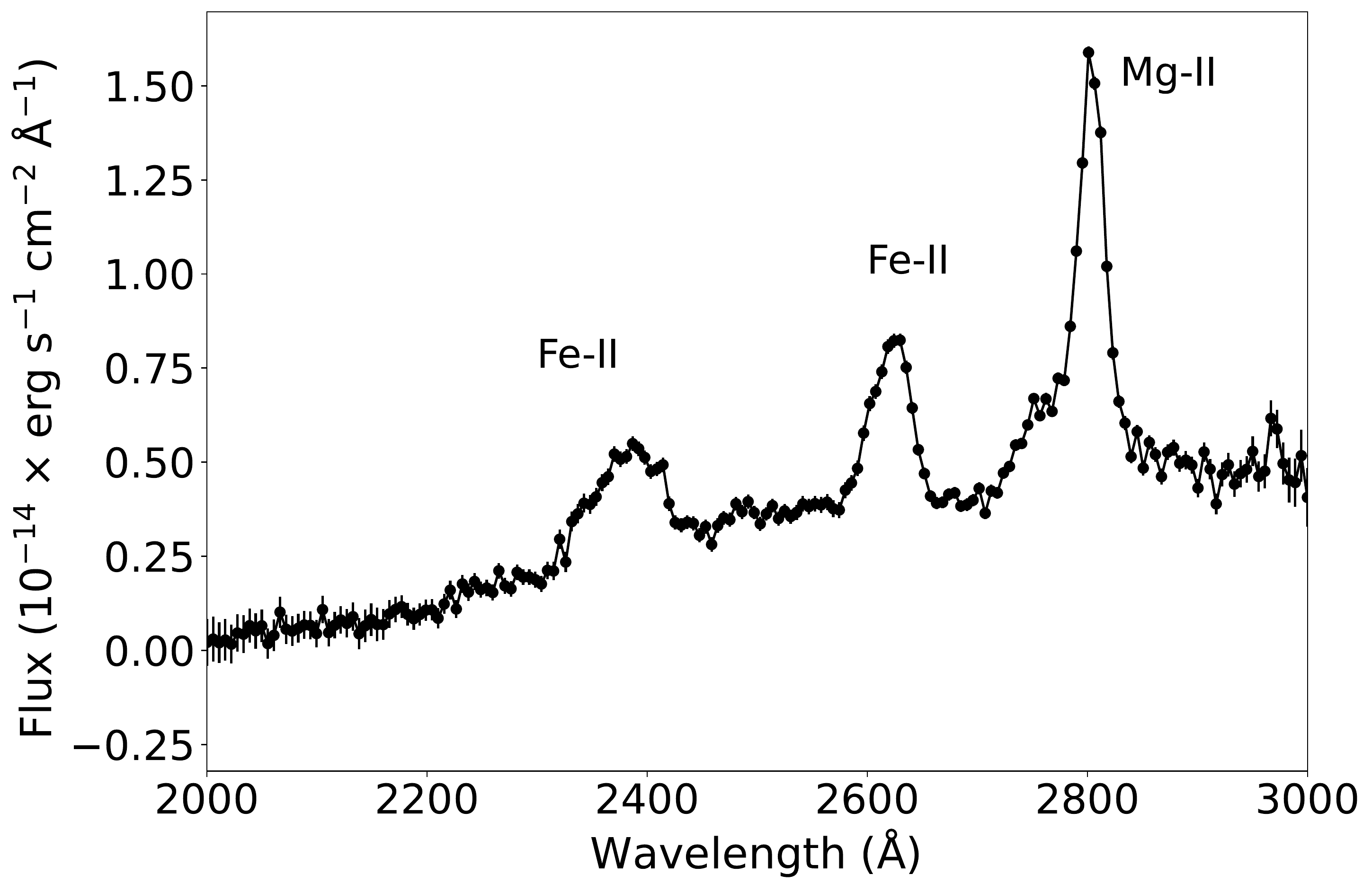}
\caption{UVIT FUV (left) and NUV (right) spectra of HIP 23309. Multiple NUV lines are detected; detections of FUV lines are much weaker, with only C {IV} being clearly detected. \label{lines_uvit}}
\end{figure}

\subsection{Robustness of UVIT M-dwarf Spectra}
We next assess the accuracy of our spectra of HIP 23309 from UVIT. For the far-UV wavelength regime, we are able to directly compare our UVIT observations to contemporaneous FUV measurements collected by HST within 1 day of this work. (\autoref{tab:hip23309_obs}). For the near-UV data, we evaluate the orbit-by-orbit stability of our measured spectrum.

\subsubsection{FUV Spectrum: Comparison to HST Data}
We coordinated our UVIT observations of HIP 23309 with HST STIS observations of HIP 23309 collected as part of the FUMES survey (PI: J. S. Pineda; \citealt{Pineda2021, Youngblood2021}). The spectral coverage of the UVIT FUV Grating1 and the HST STIS overlap from 130-170 nm, enabling us to compare spectra derived from the two facilities. 

 \begin{deluxetable}{lcccccr}
\tablecaption{Summary of HIP 23309 Observations Discussed in this Work\label{tab:hip23309_obs}}
\tablewidth{0pt}
\tablehead{
\colhead{Facility} & \colhead{Instrument} & \colhead{Grism/Grating} & \colhead{Orbits} & \colhead{Observation Start Time (UT)} & \colhead{Total Integration Time} & \colhead{Reference}
 }
\startdata
HST  & STIS & G140L (115-173 nm) & 1 & 2017-11-24 01:15:22 & 2350 s &  \citet{Pineda2021} \\
AstroSat  & UVIT & FUV Grating1 (130-180 nm)\tablenotemark{a} & 7 & 2017-11-24 10:16:44 & 11967 s &  This Work \\ 
AstroSat  & UVIT & NUV (200-300 nm)\tablenotemark{a} & 7 & 2017-11-24 10:16:44 & 12027 s &  This Work \\
\enddata
\tablenotetext{a}{\citet{Tandon2017}}
\end{deluxetable}

The UVIT FUV and STIS measurements of HIP~23309 broadly agree to within a factor of $\sim2$, but the UVIT measurements systematically exceed the STIS measurements (Figure~\ref{fig:hst_uvit_fuv}). The 130-170 nm continuum flux measured by UVIT is $(5.62 \pm 0.34)\times10^{-13}$ erg s$^{-1}$ cm$^{-2}$, compared to  $(4.02 \pm 0.04)\times10^{-13}$ erg s$^{-1}$ cm$^{-2}$. This corresponds to a 40\% overestimate by UVIT and 4.63$\sigma$ tension relative to STIS. To compare the STIS data to the UVIT data spectrally, we (1) convolve the STIS data against a Gaussian with full-width at half-max of 14.63, corresponding to the resolution of the UVIT FUV grating \citep{Dewangan2021}, and (2) rebin the smoothed spectrum to the same wavelength scale as the UVIT data using the \texttt{coronograph} package and conserving total SNR in each bin \citep{Robinson2016, Lustig2019coronagraph}. The synthetic spectrum derived from the HST data is in tension with the UVIT spectrum by an average of 1.46$\sigma$ (Figure~\ref{fig:hst_uvit_fuv}). The disagreement is strongest in the continuum region between lines, where HIP 23309 emits very little flux ($3.10\pm0.09\times10^{-16}$ erg s$^{-1}$ cm$^{-2}$ \AA$^{-1}$ in FUV; \citealt{Pineda2021}). For example, integrated from 1410-1510 \AA, the flux measured by UVIT is $5.5\pm1\times10^{-14}$ erg cm$^{-2}$ s$^{-1}$ \AA$^{-1}$, whereas the flux measured by HST is $3.39\pm0.1\times10^{-14}$ erg cm$^{-2}$ s$^{-1}$ \AA$^{-1}$, a tension of 1.9$\sigma$. The agreement between HST and UVIT is better for major lines, i.e. C\textsc{iv}. The C\textsc{iv} line emission from UVIT is $9.8\pm 3.5 \times10^{-14}$ erg cm$^{-2}$ s$^{-1}$, compared to $6.9\pm0.2\times10^{-14}$ erg cm$^{-2}$ s$^{-1}$ from HST \citep{Pineda2021}. We conclude that the UVIT FUV data broadly reproduce the HST data to within a factor of 2, but that there is the potential for uncorrected systematic error in the UVIT data, including potentially an offset of $\sim2\times10^{-16}$ erg s$^{-1}$ cm$^{-2}$ \AA$^{-1}$. 

\begin{figure}[ht!]
\plotone{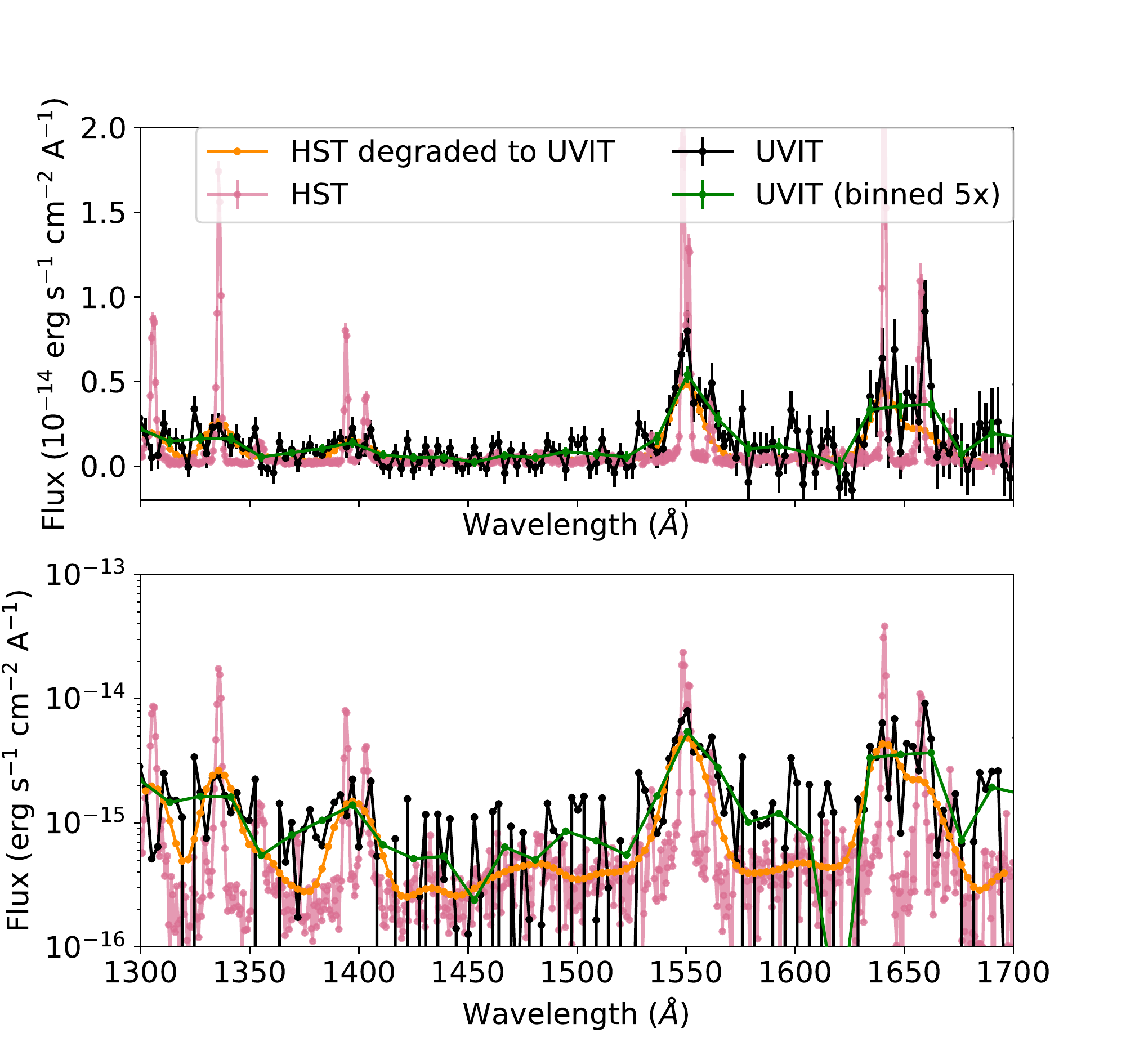} 
\caption{Comparison between HST and AstroSat-measured SEDs of HIP 23309 in the FUV (1300-1700 A). Top plot shows data in linear y-scale with error bars; bottom plot is identical to top plot except with log y-scale and omission of error bars for easier viewing. The cyan and black data correspond to the unbinned HST and UVIT data, respectively. The blue points correspond to HST data degraded to the approximate quality of the AstroSat data, to enable a closer comparison. The green points correspond to the AstroSat data binned down by a factor of 5, to improve SNR and enable better visual comparison. We assumed Gaussian uncorrelated noise when rebinning the AstroSat data. \label{fig:hst_uvit_fuv}}
\end{figure}

\subsubsection{NUV Spectrum: Stability of Measurement}
The SNR of the NUV data is high enough to support orbit-by-orbit decomposition (Figure~\ref{fig:diff_orbit}). The band-integrated (220-290 nm) orbit-by-orbit NUV data are consistent with the mean to within 10\% but vary from it by as much as $8.4\sigma$. We first consider potential systematic effects within our data before discussing physical interpretations.

It is possible that the measurement errors are underestimated, e.g. due to an uncharacterized source of red noise. Indeed, the orbit-by-orbit band-integrated fluxes display a weak increase with orbit number, {a potential systematic effect with the observatory}. Comparing the orbit-by-orbit data to the mean on a spectral basis, we find that the average deviation from the mean $A$:

\begin{equation}
    A=\frac{1}{N}\sum_{i=0}^{N}\frac{F_i-\bar{F_i}}{\sqrt{\sigma_{i}^2+\bar{\sigma_{i}}^2}},
\end{equation}

\noindent has values ranging from $1.83$ to $3.74$. This indicates that the errors are could be underestimated by a factor of a few. 

Alternately, our measurements may reflect variations in stellar flux due to activity \citep{Loyd2014, Loyd2018ApJ...867...70L,Duvvuri2023}. We do not detect evidence of large flares (variation in total flux of $\geq2\times$) on minute timescales. Stochastic variability on short timescales in the Ca~\textsc{ii} K line for HIP~23309 is on the order of 5\% \citep{Duvvuri2023}, simultaneous with our FUV data set, several hours prior to the start of our NUV data. \citet{Loyd2018ApJ...867...70L} also demonstrate $\sim$10\% broadband quiescent FUV variability in their young M-dwarf sample, with  one object displaying up to 50\% variations relative to the median, which can translate to factors of a few from minimum to maximum. Our observed NUV variability is consistent with these data as we expect the NUV variability to fall between the optical and FUV measurements. We also note that the fractional stochastic variability may not be correlated with absolute activity \citep{Duvvuri2023}.

However, our NUV data show a clear trend over time (Figure~\ref{fig:diff_orbit}. This might be explained through rotational variation as an active region comes into view, enhancing the total NUV flux. To test this idea, we fit a simple sinusoidal model ($f(t) = f_{0} + A\cos(\phi + \omega t)$) to the 7 NUV data points, with $\omega = 2\pi/P$ for $P = 8.6 \pm 0.07$~d. We find that the best solution places our data points at the rising part of the sinusoidal trough, with 50-70\% fractional variation about the model median total NUV flux. This result appears to be similar to that of the most quiescent variable target in \citet{Loyd2018ApJ...867...70L}. Although, we cannot confirm an astrophysical origin for the observed NUV variability, it is plausible.

\begin{figure}[ht!]
\plottwo{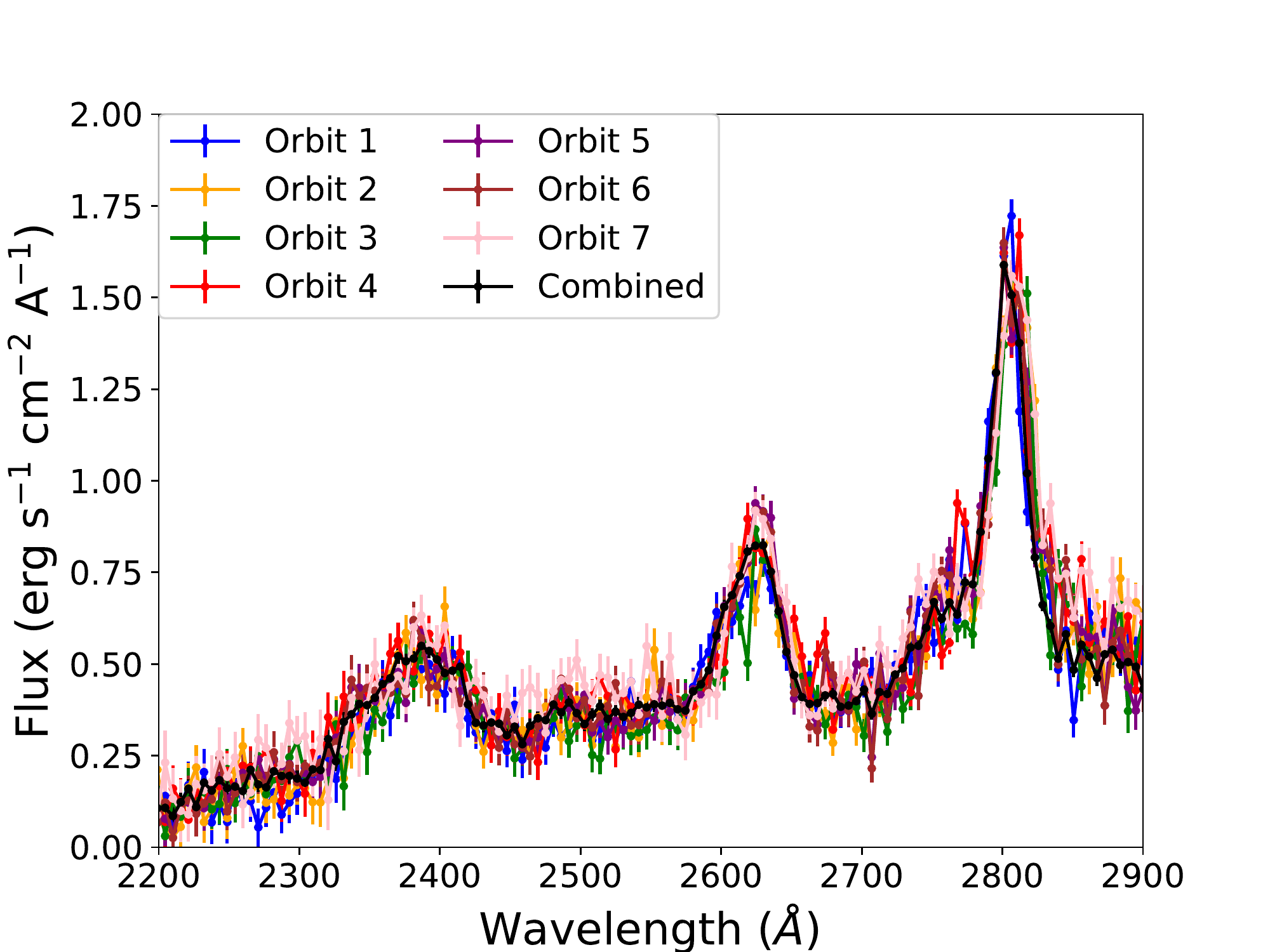}{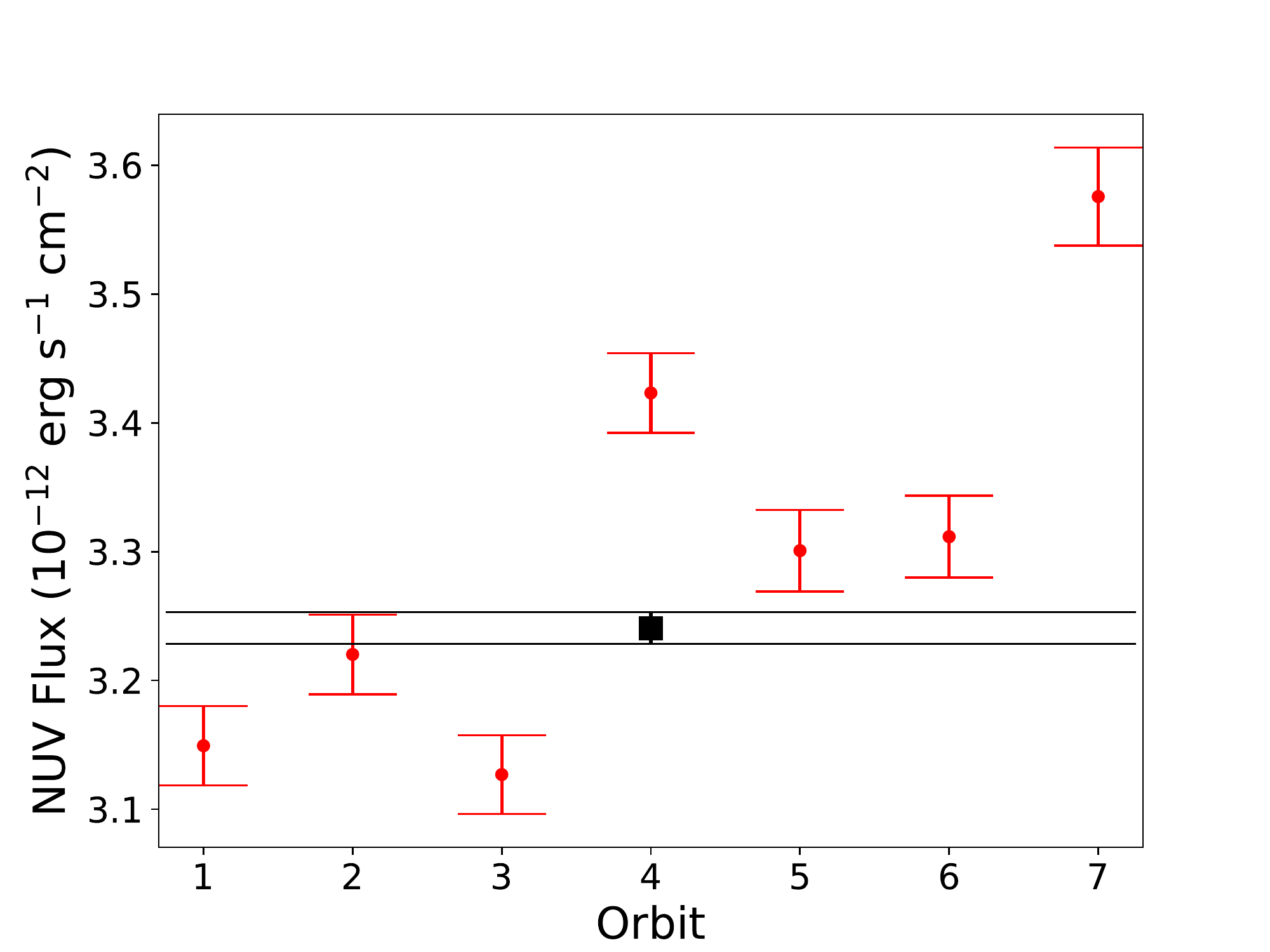}
\caption{Orbit-by-orbit NUV spectral (left) and band-integrated (right) flux of HIP 23309 as measured by UVIT. The band-integrated orbit-by-orbit fluxes agree with the mean flux to within 10\% but are in statistically significant tension with the mean. This indicates that either our errors are underestimated, or that there is stellar variability at the $\sim10\%$ level. \label{fig:diff_orbit}}
\end{figure}

\subsection{HIP 23309 SED in Context}
We compare our FUV and NUV spectra of HIP 23309 to spectra of other low-mass stars to assess whether our measurements are sensible in context (Figure~\ref{fig:luminosity_comparison}). We specifically compare the spectral flux at the stellar surface of HIP 23309 to MUSCLES M-dwarf and K-dwarf spectra (v2.2, \citealt{France2016, Youngblood2016, Loyd2016}), to the composite VPL spectrum of AD Leo \citep{Segura2005} and to StarCAT measurements of AU Mic in the FUV \citep{Ayres2010}. The MUSCLES targets chosen are mature stars which bracket HIP 23309 in spectral type, whereas AD Leo is a highly active M-dwarf and AU Mic is a young pre-main sequence (22 Myr, \citealt{Mamajek2014, Plavchan2020}) early M-dwarf star similar to HIP 23309. HIP 23309's FUV emission is similar to AU Mic's, which is sensible in that AU Mic is of similar spectral type and age to HIP 23309. HIP 23309 is brighter in the FUV and NUV compared to main-sequence M-dwarfs of similar spectral type, consistent with the finding that pre-main sequence M-dwarfs display elevated UV emission compared to main-sequence M-dwarfs due to enhanced stellar activity and larger radii \citep{Shkolnik2014}. This additional activity flattens the slope of the spectrum across the entire NUV band, as seen in the most active M-dwarfs \citep{Segura2005, Walkowicz2008}, with much stronger emissions at short wavelengths relative to the main-sequence M-dwarfs. At longer wavelengths in the NUV, the HIP 23309 stellar flux begins to converge with main-sequence stars of similar spectral type, which is consistent with the increasing importance of photospheric emission at longer wavelengths. We conclude that to zeroth order, our measured NUV and FUV spectra of HIP 23309 are sensible in the context of other measured spectra of low-mass stars. 

\begin{figure}[ht!]
\plottwo{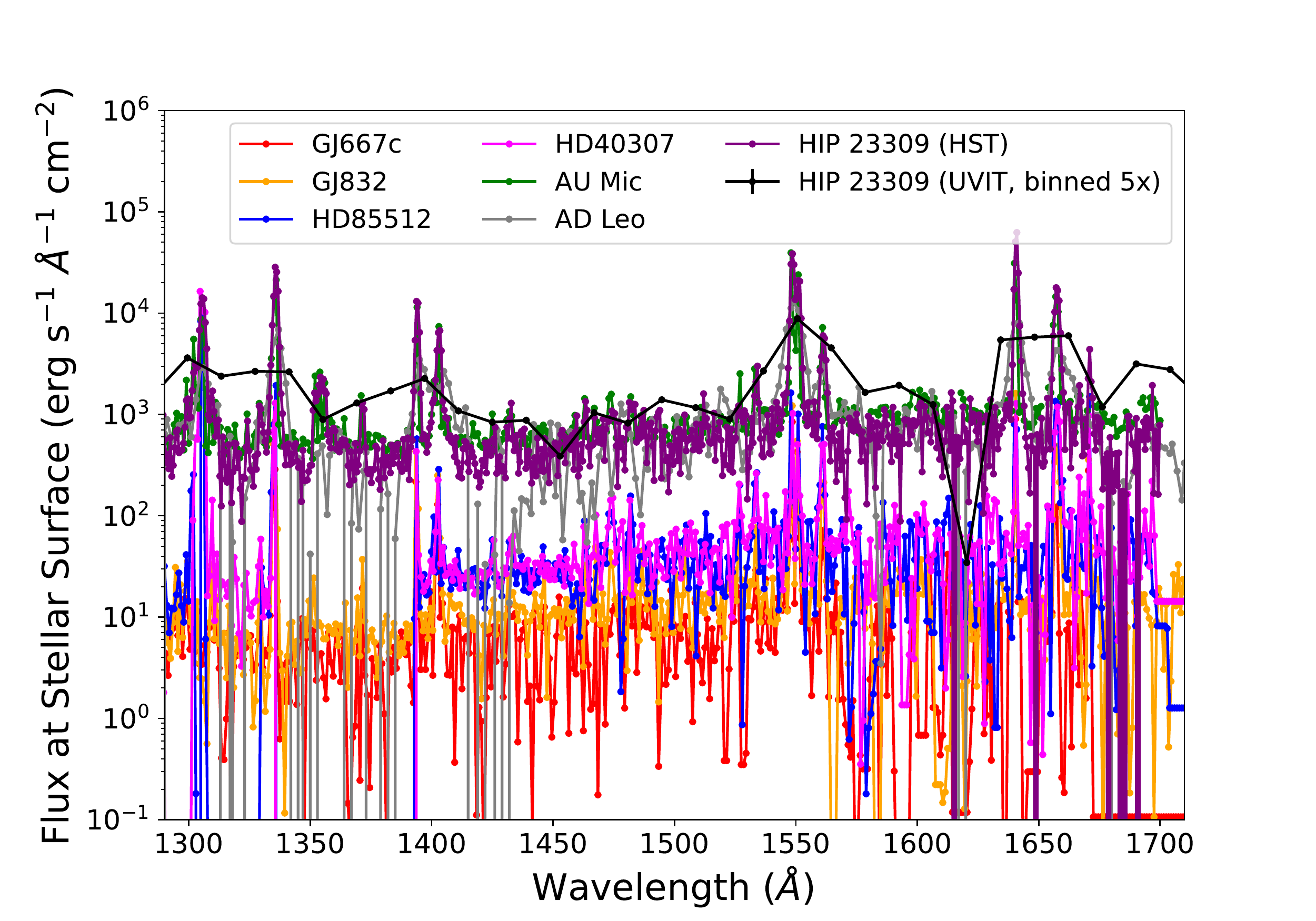}{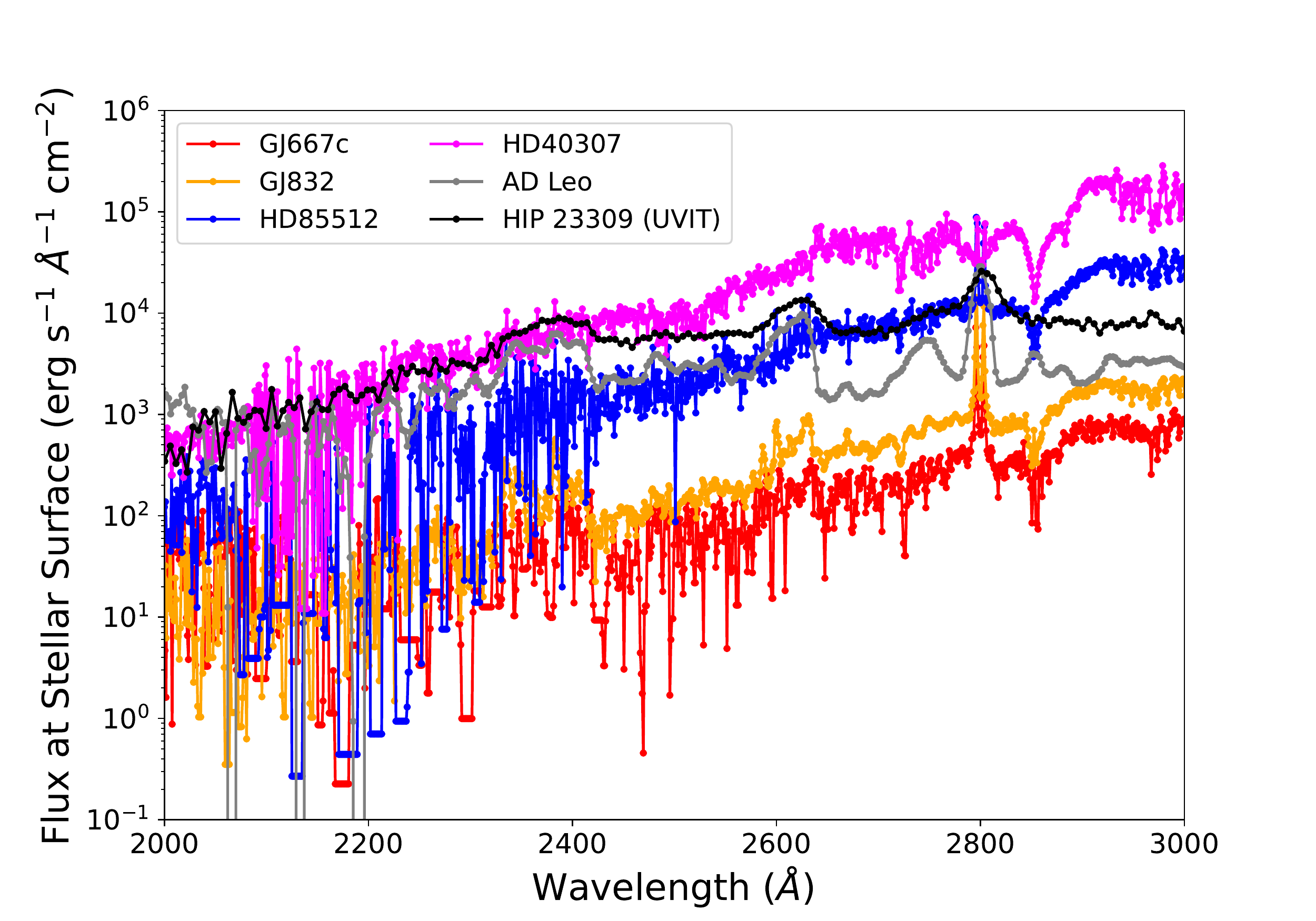}
\caption{Stellar flux of HIP 23309 at the stellar surface ($F_\star\times\frac{d_\star^2}{r_\star^2}$) in the FUV (left) and NUV (right), compared to other low-mass stars. Main-sequence comparison stars are ordered by $T_{eff}$ \citep{Loyd2016}. HIP 23309's stellar flux is much higher than mature stars of similar spectral type at short wavelengths due to enhanced activity because of its young age but converges with them at longer wavelengths where photospheric emission dominates. HIP 23309's flux in the FUV is similar to AU Mic's, another extremely young M-dwarf of a similar spectral class. Overall, our spectra of HIP 23309 appear sensible in the context of other measured spectra of low-mass stars. \label{fig:luminosity_comparison}}
\end{figure}


\section{Applications of UV SEDs to Rocky Exoplanet Modeling}
In this section, we explore the implications of our measured SED of HIP 23309 for a hypothetical temperate terrestrial planet orbiting HIP 23309. Our intent is to illustrate the utility of M-dwarf SEDs for studying rocky exoplanets, and particularly for questions of astrobiological interest.

\subsection{Implications of HIP 23309 SED for UV-Dependent Prebiotic Chemistry\label{sec:prebio}}
UV light has been proposed as necessary to the origin of life, but the surfaces of habitable M-dwarf planets are low-UV environments \citep{Ranjan2017c}. \citet{Rimmer2018} showed that HCN photohomologation chemistry used in proposed prebiotic pathways for ribonucleotide synthesis \citep{Xu2018, Xu2020, Rimmer2021} require $F_{200-280nm}\geq(6.8\pm 3.6)\times10^{10}$ cm$^{-2}$ s$^{-1}$ nm$^{-1}$ to function, where $F_{200-280nm}$ is the surface flux averaged from 200-280 nm. This means that if HCN photohomologation is required for the emergence of life, then life cannot emerge on planets that have access to less than this critical flux of UV. \citet{Rimmer2018} find that the quiescent emission from known M-dwarfs is insufficient to meet this criterion and that HCN photohomologation chemistry can only proceed on planets orbiting M-dwarfs if enabled by transiently enhanced UV from flares. Laboratory efforts to experimentally determine if flare UV can substitute for quiescent UV in HCN photohomologation are now under way\footnote{\url{https://www.mrao.cam.ac.uk/~pbr27/resources.html}, accessed 12/7/2022}. If flares cannot compensate for low-steady state UV, then life search on M-dwarf exoplanets may enable tests of theories the origin of life \citep{RimmerRanjan2021, Rimmer2023}.

We consider whether the UV emission from HIP 23309 meets the \citet{Rimmer2018} criterion. We assembled an FUV-NIR SED of HIP 23309 by concatenating our UVIT-measured FUV and NUV fluxes to the low-resolution GAIA spectrum of HIP 23309 \citep{GaiaCollaboration2022}. We set the stellar flux at wavelengths shorter than the limits of the measured SED to 0. We scale these observed fluxes to the flux that would be experienced by a hypothetical Earth-like planet orbiting HIP 23309 by multiplying them by $(\frac{d_{*}}{a})^2$, where $d_{*}=26.8$ pc is the distance to HIP 23309 \citep{Malo2014, GaiaCollaboration2022}, $a=1\texttt{AU}\times(\frac{L_{*}}{L_\sun*0.9})^{0.5}$ is the distance such that the bolometric flux received at distance $a$ is equivalent to the bolometric flux received by Earth, corrected by $0.9\times$ to account for the redder emission from M-dwarf stars \citep{Segura2005}, and $L_{*}=6.82\times10^{32}$ erg s$^{-1}$ \citep{Pineda2021}. We then calculated the hemispherically-integrated surface actinic flux (``surface radiance") for an early Earth-like planet with a clear-sky 0.9 bar N$_2$, 0.1 bar CO$_2$ atmosphere as in \citet{Ranjan2017c}. We find that HIP 23309 comfortably meets the \citet{Rimmer2018} criterion, with $F_{200-280nm}=3\times10^{11}$ cm$^{-2}$ s$^{-1}$ nm$^{-1}$. Interestingly, we also find that the quiescent flux from the extremely active M-dwarf AD Leo also meets the \citet{Rimmer2018} criterion ($F_{200-280nm}=2\times10^{11}$ cm$^{-2}$ s$^{-1}$ nm$^{-1}$), and that quiescent flux from early (high-mass) M-dwarfs like GJ 832 also marginally meet this criterion. This contradiction with \citet{Rimmer2018} is due to their assignment of a ceiling on single-scattering albedo $\omega_0$ of 0.9\footnote{P. Rimmer, personal communication, 12/7/2020} when calculating two-stream radiative transfer, which underestimates surface UV in the unique regime unveiled on anoxic early Earth-like planets \citep{Ranjan2017a, Rimmer2021}. We instead find that planets orbiting young or active M-dwarf stars, as well as some quiescent early M-dwarf stars, can meet the \citet{Rimmer2018} criterion, illustrating the necessity of measuring and constraining the evolution of NUV SEDs of individual planet-hosting M-dwarfs to understand the habitability of their worlds \citep{RimmerRanjan2021}. 

\begin{figure}[ht!]
\plotone{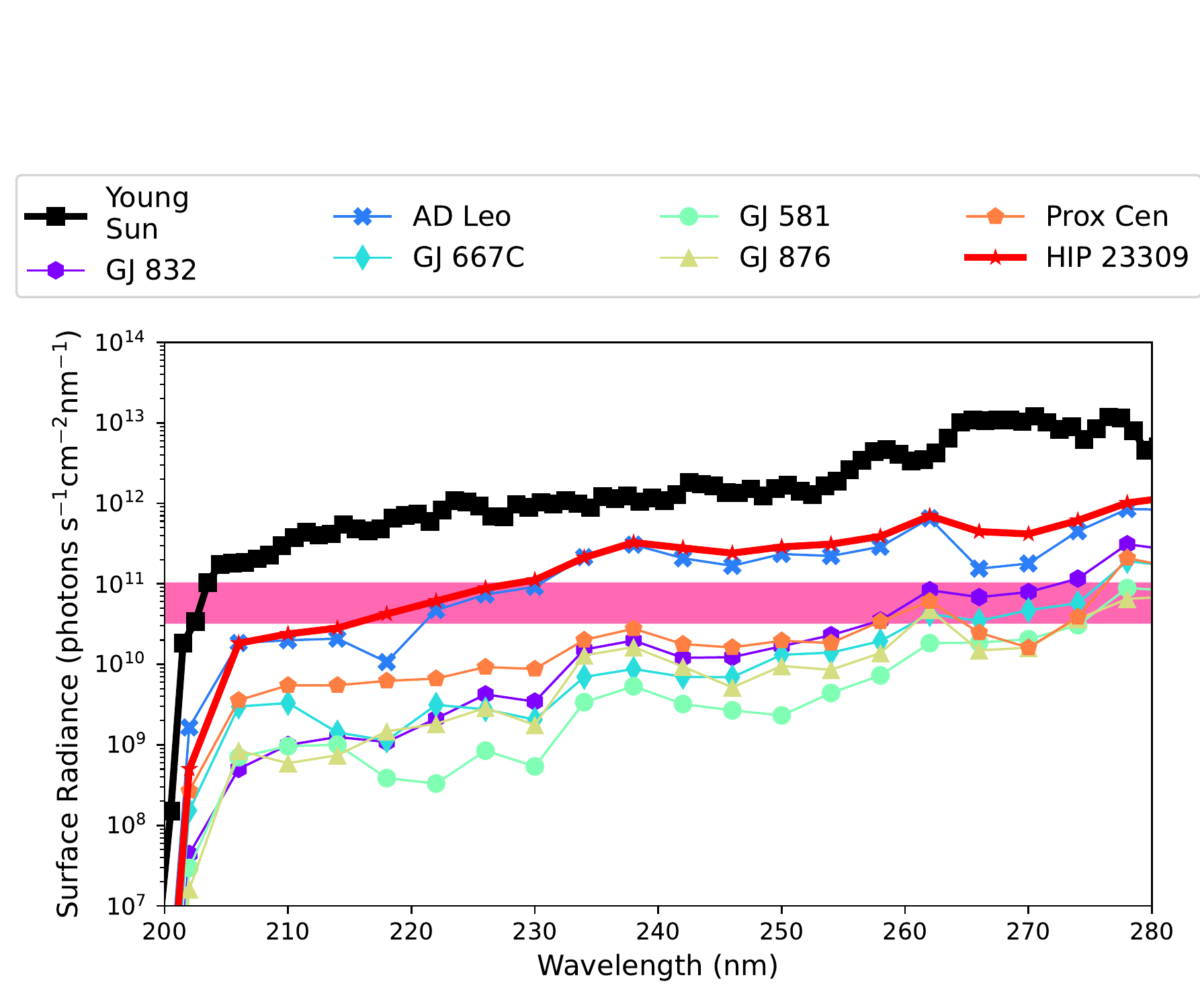}
\caption{Surface radiances on early Earth-like planets orbiting different M-dwarfs including HIP 23309, computed as in \citet{Ranjan2017c}. The \citet{Rimmer2018} minimum flux criterion for HCN homologation towards ribonucleotide synthesis is shown in pink. Extremely active stars like HIP 23309 meet the \citet{Rimmer2018} criterion, illustrating the need to constrain UV emission of M-dwarfs over the course of their lifetimes to understand the potential of life search on M-dwarf planets to test theories of abiogenesis \citep{RimmerRanjan2021}. \label{fig:prebioche}}
\end{figure}

\subsection{Implications of HIP 23309 SED for Atmospheric Photochemistry}
In this section, we explore the utility and limitations of AstroSat UVIT for constraining the UV SEDs of exoplanet host stars for atmospheric photochemistry applications. UV photochemistry controls the trace gas composition of temperate terrestrial planet atmospheres, and accurate UV SEDs are necessary for accurate predictions of atmospheric composition and inferences derived thereof \citep{Rugheimer2015mdwarf, Teal2022}. We assess the impact of UV SEDs on predicted atmospheric trace gas concentrations using a procedure established first by \citet{Kasting1997} and subsequently extensively deployed (e.g., \citealt{Segura2005, Rauer2011, Teal2022}): We run a photochemical model with a fixed planetary scenario, varying only the UV irradiation between model runs, and examine the implications for predicted atmospheric composition. Our calculations are solely sensitivity tests aimed  at constraining the utility and limitations of using AstroSat UVIT to constrain M-dwarf SEDs for photochemical modeling applications, as opposed to self-consistent predictions of the atmospheric composition of temperate terrestrial planets orbiting young M-dwarf stars. 

To conduct this sensitivity test, we use the MIT Exoplanet Atmospheric Chemistry model (MEAC, \citealt{Hu2012}), as updated by \citet{Ranjan2022b}. We apply MEAC to the N$_2$-dominated abiotic temperate terrestrial planet benchmark scenario from \citet{Hu2012}, which simulates a planet geologically similar to Earth in the absence of life. We explore the effects of different estimates of the HIP 23309 SED on the abundances of key HCO species which dominate the atmosphere and whose remote detection could constrain surfaces processes like the carbon cycle and the presence or absence of life \citep{Bean2017, Schwieterman2019, Wogan2020,Lehmer2020, Thompson2022}.

We begin by showing that the UVIT-measured SED alone can achieve order-of-magnitude accuracy at best in the photochemical modeling of rocky exoplanet atmospheres. We formulate a UVIT-only input SED for HIP 23309 as in Section~\ref{sec:prebio}, except that we do not include the luminosity correction factor of 0.9 to maintain consistency with similar past sensitivity tests conducted with MEAC (e.g., \citealt{Seager2013b}). We formulate a second SED as the UVIT SED, except that we replace the UVIT FUV measurements with the FUMES HST measurements (Figure~\ref{fig:inputseds}).  Calculating the trace gas composition under irradiation by each of these SEDs (Figure~\ref{fig:photochem}), we find that the use of the UVIT-only SED leads to overpredictions of CH$_4$ by $7\times$ and underprediction of CO by $6\times$ relative to the UVIT+HST SED (Table~\ref{tab:photochem}). Two factors drive model inaccuracy when using the UVIT-only SED. First, the UVIT SED systematically overestimates the continuum flux due to its lower sensitivity. This leads to overestimates of H$_2$O photolysis and accounts for the variation in pCO and some of the variation in pCH$_4$. The effects of this overestimate are amplified by the large wavelength gap between reliable measurements in the FUV and NUV gratings, which is interpolated in the model. The importance of the UV continuum has been affirmed in other studies \citep{Teal2022, Peacock2022}.  Second, the UVIT SED does not reliably cover the Lyman-alpha line, where M-dwarfs emit a substantial fraction of their flux and which is particularly relevant to CH$_4$ photolysis in anoxic atmospheres \citep{France2013, Ranjan2022b}. \citet{Peacock2022} reported the predicted atmospheric composition of M-dwarf planets to be insensitive to the degree of self-reversal in the stellar Ly$\alpha$ line, and we reproduced this finding with our model. However, omitting Ly$\alpha$ entirely results in a much bigger change in irradiating flux than variations in the degree of self-reversal in the Ly$\alpha$ line, and we find this omission to result in significant variations in predicted pCH$_4$. 

\begin{figure}[ht!]
\plotone{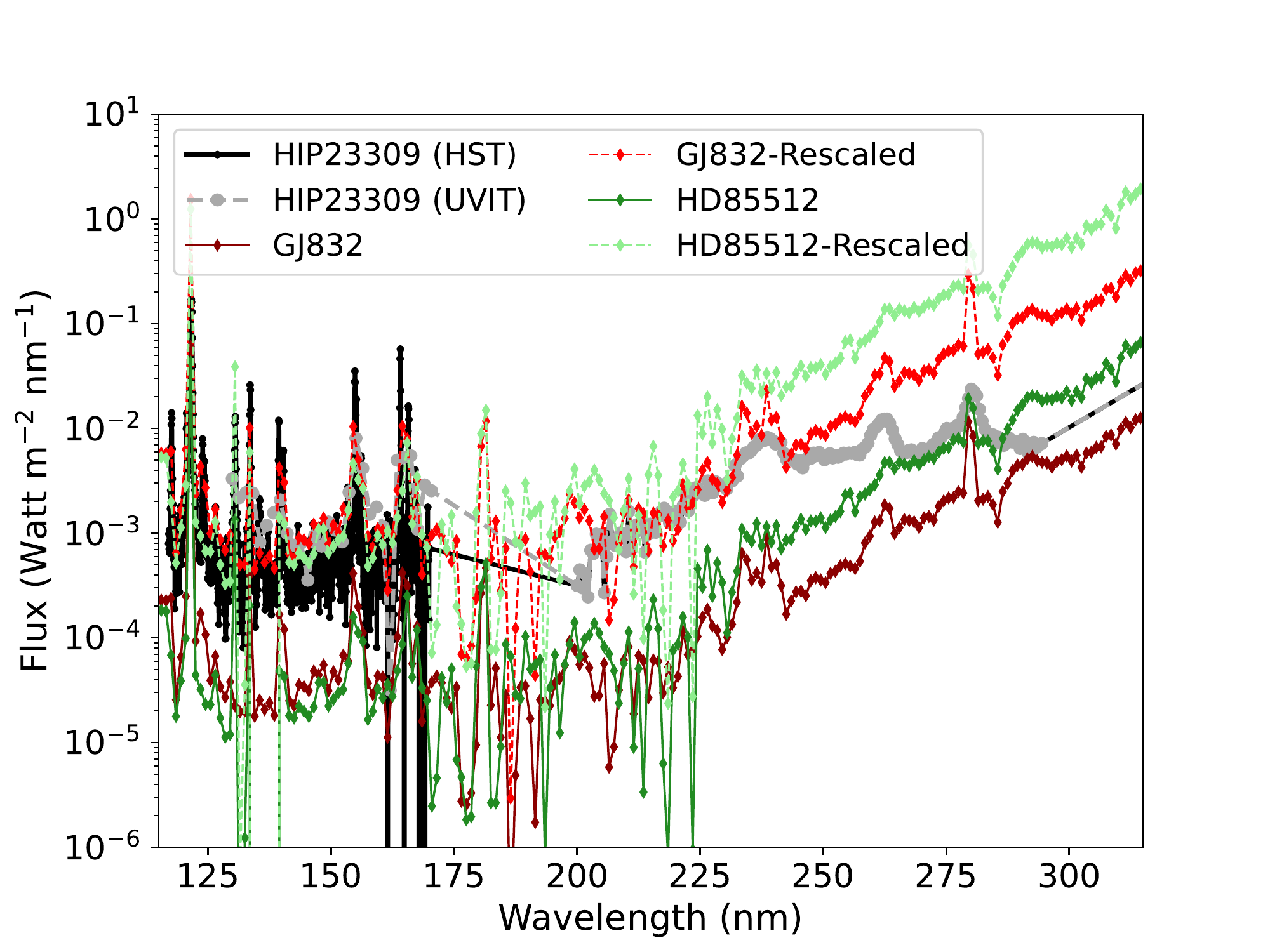}
\caption{Input UV SEDs used for modeling the atmospheric composition (Figure~\ref{fig:photochem}) of a hypothetical abiotic Earth-like planet orbiting HIP 23309.\label{fig:inputseds}}
\end{figure}

\begin{figure}[ht!]
\plotone{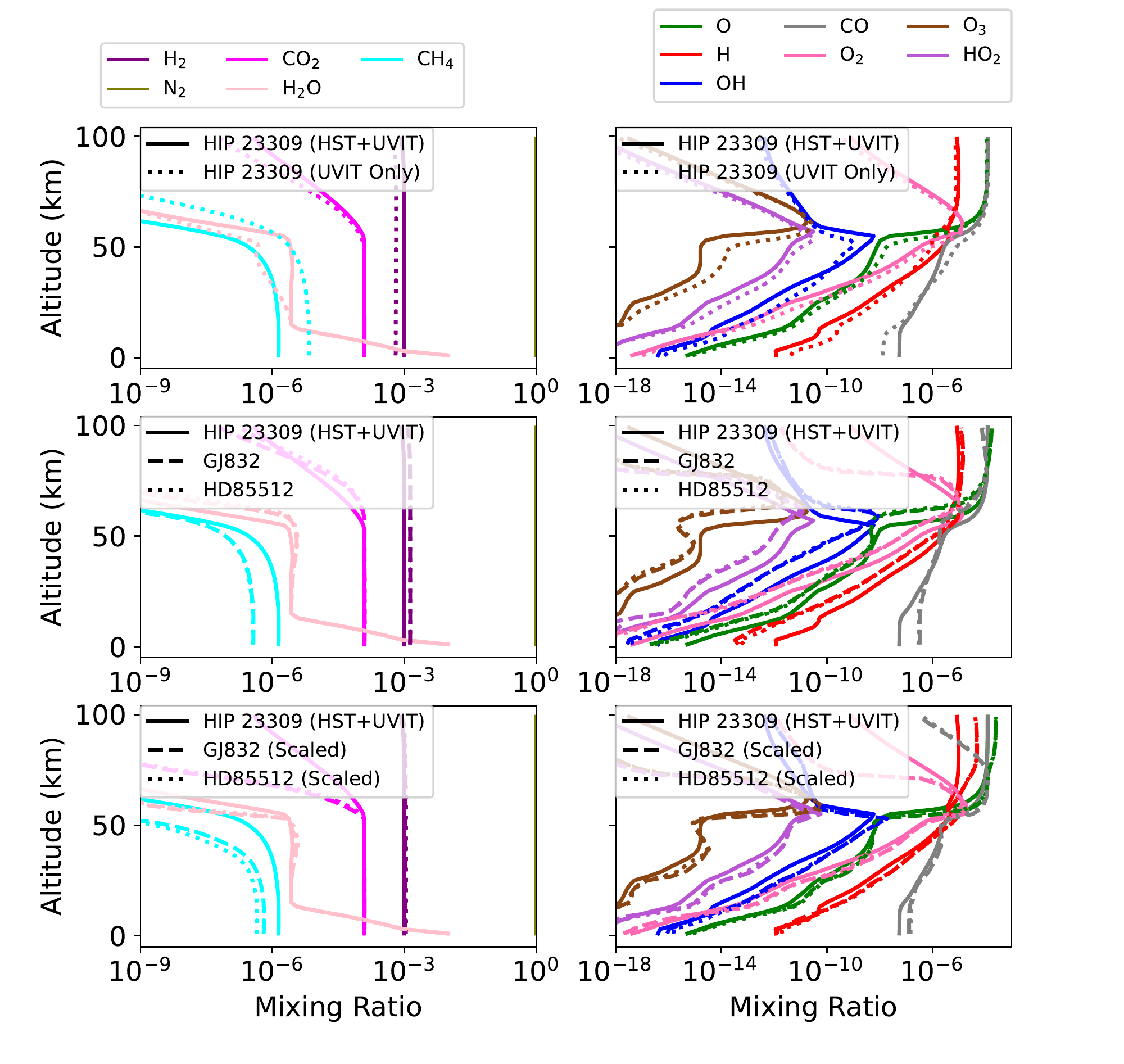}
\caption{Predicted atmospheric composition of a hypothetical abiotic Earth-like planet orbiting HIP 23309, subject to varying estimates of its UV irradiation (Figure~\ref{fig:inputseds}).The different colors correspond to different chemical species. The left column corresponds to chemical species emitted by the surface, while the right column corresponds to photochemically generated species. In each row, the different linestyles correspond to different SEDs. The UVIT-only SED alone is insufficient to achieve highly precise photochemical modeling (top row), but can be used to calibrate other estimates of the stellar SED (middle row) to achieve higher precision, as done crudely here (bottom row) and with vastly more sophistication by, e.g., \citep{Peacock2019a, Peacock2019b}))\label{fig:photochem}}
\end{figure}

\begin{deluxetable}{lcr}
\tablecaption{Surface Mixing Ratios Predicted in MEAC Abiotic N$_2$-dominated Benchmark Scenario\label{tab:photochem}}
\tablewidth{0pt}
\tablehead{
\colhead{SED} & \colhead{$r_{CH_{4}}(z=0)$} & \colhead{$r_{C0}(z=0)$}
 }
\startdata
HIP 23309 HST FUV & $1\times10^{-6}$ & $6\times10^{-8}$\\
HIP 23309 UVIT FUV & $7\times10^{-6}$ & $1\times10^{-8}$\\
GJ 832 (MUSCLES) & $4\times10^{-7}$ & $3\times10^{-7}$\\
HD85512 (MUSCLES) & $4\times10^{-7}$ & $3\times10^{-7}$\\
GJ 832 (Scaled) & $6\times10^{-7}$ & $1\times10^{-7}$\\
HD85512 (Scaled) & $4\times10^{-7}$ & $1\times10^{-7}$\\
\enddata
\end{deluxetable}

We next consider how the UVIT-only SED performs in comparison to proxy SEDs from stars of similar spectral type, which are sometimes employed when accurate SEDs of planet-hosting stars are not available (e.g., \citealt{Hu2020, Hu2021waterworlds}). We consider SEDs from GJ832 (M1.5) and HD 85512 (K6), which are well-characterized by the MUSCLES survey (v2.2, \citealt{France2016, Loyd2016, Youngblood2016}), as proxies for HIP 23309 (M0). We normalize their SEDs to the solar constant as for HIP 23309, and apply these SEDs to our planetary scenario (Figure~\ref{fig:inputseds}). We find that use of proxy SEDs matches the HST+UVIT SED calculation better than the UVIT-only SED (Figure~\ref{fig:photochem}, Table~\ref{tab:photochem}). 

We finally consider whether UVIT data can be used to calibrate other techniques for estimating M-dwarf UV SEDs. For example, partial SED measurements have been used to calibrate full-SED estimates derived from stellar models \citep{Peacock2019a, Peacock2019b, Peacock2020}. Such sophisticated estimates are beyond the scope of this work; as proof-of-concept, we instead utilize our UVIT measurements to crudely calibrate our use of similar stellar spectra as proxies. Specifically, we scale our GJ832 and HD85512 SEDs by a constant factor, such that they have the same flux integrated from 130-170 nm as our UVIT-only HIP 23309 SED (Figure~\ref{fig:inputseds}). We do not utilize the NUV data because the UVIT NUV grating is no longer operative, and so it is of limited utility to determine its utility at calibrating other estimates. For simplicity we scale the SEDs across all wavelengths, but only the UV wavelengths have any impact on our models since our photochemistry-only sensitivity test neglects climate adjustments. We find that use of scaled proxy SEDs to marginally improve agreement with the HIP 23309 UVIT+HST SED compared to unscaled proxy SEDs, in the case of GJ832 reproducing the predicted surface concentrations of CO and CH$_4$ to within a factor of 2 (Figure~\ref{fig:photochem}, Table~\ref{tab:photochem}). That even an extremely crude calibration with UVIT can marginally improve agreement suggests that it is possible that more sophisticated calibrations with UVIT can significantly improve agreement; we advocate such work.  

\section{Discussion \& Conclusions}
Constraining the UV SEDs of low-mass stars is crucial to interpreting atmospheric measurements of the exoplanets they host, and to understanding the habitability and urability of those planets \citep{Deamer2022}. The primary instrument used for such studies, HST, is oversubscribed and well past its designed lifetime. Here, we have examined the potential of  AstroSat UVIT to carry out such studies through spectrally resolved NUV and FUV grating observations of the young M0 star HIP 23309. We robustly detect stellar lines of C\textsc{iv} in the FUV and Fe\textsc{ii} and Mg\textsc{ii} in the NUV. We find UVIT can reproduce contemporaneous HST measurements in the FUV to within a factor of 2 and that its orbit-by-orbit NUV measurements are stable to 10\%, but that its errors may be underestimated by a factor of a few. The UVIT-measured HIP 23309 SED displays trends in the context of other measured M-dwarf SEDs that are consistent with zeroth-order expectations. Taken together, the example of HIP 23309 suggests AstroSat UVIT is a viable tool for characterizing the UV SEDs of bright, nearby low-mass stars to within a factor of a few. 

We have demonstrated the utility of such measurements for planetary applications using HIP 23309 as a case study. Archival measurements of stellar UV from the UVIT NUV grating can inform studies of planetary habitability because they constrain emission at the $\geq 190$ nm wavelength range relevant to habitable planet surfaces \citep{Ranjan2017a}. For example, HIP 23309 meets the \citet{Rimmer2018} criterion for sufficient UV output to power HCN photohomologation in its habitable zone, demonstrating the necessity of characterizing the detailed evolution of M-dwarf SEDs to understand the potential of biosignature search on M-dwarf planets to test theories of the origin of life \citep{RimmerRanjan2021, Rimmer2023}. Our FUV and NUV spectra of HIP 23309 on their own permit limited accuracy in photochemical modeling of planetary atmospheres, because the UVIT FUV grating does not reliably cover the crucial Lyman alpha line and because the limited sensitivity and wavelength coverage of the FUV grating leads to overestimates of the continuum flux. However, our sensitivity tests suggest that UVIT FUV observations can be used to calibrate estimates of M-dwarf SEDs, thereby supporting photochemical modeling. Our work suggests a role for archival NUV observations and archival and future FUV observations in supporting studies of exoplanet habitability and photochemistry. 

Our study bears implications for the proposed Indian Spectroscopic and Imaging Space Telescope \citep[INSIST;][]{Subramaniam2022}. The INSIST mission design features a 1m mirror ($\sim 7\times$ the size of UVIT), suggesting a factor of $\sim$7 improvements in signal-to-noise relative to UVIT, and continuous wavelength coverage from the FUV to the NUV (150-300 nm). This means that INSIST would likely outperform UVIT at characterizing the important UV continuum. However, the termination of its wavelength coverage at 150 nm means that INSIST would not be able to directly constrain the 121.6 nm Lyman alpha line which can be important to temperate exoplanet photochemistry. We advocate for the extension of INSIST's spectral coverage to 120 nm or shorter, to enable direct characterization of the full wavelength range relevant to exoplanet photochemistry. However, higher resolution modes ($R\sim10,000$) may be necessary to accurately measure Lyman alpha fluxes accounting for the impact of the intervening ISM \citep{Youngblood2021}. Precise measurements of chromospheric and transition region lines like Mg\textsc{ii}, C\textsc{iv}, Si\textsc{iv}, and/or N\textsc{v} can provide useful proxies within potential INSIST data to estimate the total FUV flux of interesting stellar targets \citep{Pineda2021}. Nevertheless, even if INSIST is unable to directly measure full exoplanet host star UV SEDs, measurements from INSIST will still be able to calibrate other estimates of M-dwarf SEDs like stellar models \citep{Peacock2019a, Peacock2019b}, and its coverage of NUV wavelengths will enable planetary habitability studies (e.g., \citealt{Rugheimer2015, Ranjan2017c, Rimmer2018, O'Malley-James2019}). 

\acknowledgements
JSP acknowledges support from NASA via grant HST-GO-14640. SR thanks Sujan Sengupta for his mentorship in developing the project idea, and the Indian Institute of Astrophysics, Northwestern University, and the University of Arizona for their support while carrying out this project. PKN acknowledges TIFR's postdoctoral fellowship and support from the Centro de Astrofisica y Tecnologias Afines (CATA) fellowship via grant Agencia Nacional de Investigacion y Desarrollo (ANID), BASAL FB210003. We thank Sarah Peacock and Edward Schwieterman for sharing TRAPPIST-1 spectra from \citet{Peacock2022} as an additional validation test. We thank Gulab Dewangan for sharing the raw data for HZ4 to enable us to verify correct functioning of our pipeline. We thank the anonymous referee for their constructive critical feedback, which has materially improved the quality of this manuscript.

\textit{Author contribution statement}: SR conceived the project, served as co-I on the UVIT proposal, assessed the reduced data, performed the modeling, and wrote the paper. PN led to the reduction of the UVIT data. JSP served as co-I on the UVIT proposal, advised on data reduction, supplied pointers to other M-dwarf spectra utilized in this paper, and commented on fundamental stellar physics. MN was involved with the data validation by developing an alternate reduction pipeline (Narang 2023) with PN to reduce the UVIT data. All authors reviewed and edited the paper.

\textit{Code availability}: The code underlying Sections 3.2.1-3.3 and Figures~\ref{fig:hst_uvit_fuv}, \ref{fig:diff_orbit}, \ref{fig:luminosity_comparison}, and \ref{fig:inputseds} is available at \url{https://github.com/sukritranjan/uvit_mdwarf}. The code underlying Section 4.1 and Figure~\ref{fig:prebioche} is modified from \citet{Ranjan2017c} and is available at \url{https://github.com/sukritranjan/ranjanwordsworthsasselov2017b/tree/for-uvit-paper}. The input files and simulation outputs from the MEAC code underlying Section 4.2 and Figure~\ref{fig:photochem} are available by request. 

\vspace{5mm}
\facilities{HST(STIS), AstroSat (UVIT)}
\software{MEAC \citep{Hu2012}}






\bibliography{sample63}{}

\begin{thebibliography}{}
\expandafter\ifx\csname natexlab\endcsname\relax\def\natexlab#1{#1}\fi
\providecommand{\url}[1]{\href{#1}{#1}}
\providecommand{\dodoi}[1]{doi:~\href{http://doi.org/#1}{\nolinkurl{#1}}}
\providecommand{\doeprint}[1]{\href{http://ascl.net/#1}{\nolinkurl{http://ascl.net/#1}}}
\providecommand{\doarXiv}[1]{\href{https://arxiv.org/abs/#1}{\nolinkurl{https://arxiv.org/abs/#1}}}

\bibitem[{{Ayres}(2010)}]{Ayres2010}
{Ayres}, T.~R. 2010, \apjs, 187, 149, \dodoi{10.1088/0067-0049/187/1/149}

\bibitem[{{Batalha} {et~al.}(2018){Batalha}, {Lewis}, {Line}, {Valenti}, \&
  {Stevenson}}]{Batalha2018}
{Batalha}, N.~E., {Lewis}, N.~K., {Line}, M.~R., {Valenti}, J., \& {Stevenson},
  K. 2018, \apjl, 856, L34, \dodoi{10.3847/2041-8213/aab896}

\bibitem[{{Bean} {et~al.}(2017){Bean}, {Abbot}, \& {Kempton}}]{Bean2017}
{Bean}, J.~L., {Abbot}, D.~S., \& {Kempton}, E. M.~R. 2017, \apjl, 841, L24,
  \dodoi{10.3847/2041-8213/aa738a}

\bibitem[{{Bell} {et~al.}(2015){Bell}, {Mamajek}, \& {Naylor}}]{Bell2015}
{Bell}, C. P.~M., {Mamajek}, E.~E., \& {Naylor}, T. 2015, \mnras, 454, 593,
  \dodoi{10.1093/mnras/stv1981}

\bibitem[{{Blanco} {et~al.}(1979){Blanco}, {Catalano}, \&
  {Marilli}}]{1979Natur.280..661B}
{Blanco}, C., {Catalano}, S., \& {Marilli}, E. 1979, \nat, 280, 661,
  \dodoi{10.1038/280661a0}

\bibitem[{{Bryson} {et~al.}(2021){Bryson}, {Kunimoto}, {Kopparapu}, {Coughlin},
  {Borucki}, {Koch}, {Aguirre}, {Allen}, {Barentsen}, {Batalha}, {Berger},
  {Boss}, {Buchhave}, {Burke}, {Caldwell}, {Campbell}, {Catanzarite},
  {Chandrasekaran}, {Chaplin}, {Christiansen}, {Christensen-Dalsgaard},
  {Ciardi}, {Clarke}, {Cochran}, {Dotson}, {Doyle}, {Duarte}, {Dunham},
  {Dupree}, {Endl}, {Fanson}, {Ford}, {Fujieh}, {Gautier}, {Geary},
  {Gilliland}, {Girouard}, {Gould}, {Haas}, {Henze}, {Holman}, {Howard},
  {Howell}, {Huber}, {Hunter}, {Jenkins}, {Kjeldsen}, {Kolodziejczak},
  {Larson}, {Latham}, {Li}, {Mathur}, {Meibom}, {Middour}, {Morris}, {Morton},
  {Mullally}, {Mullally}, {Pletcher}, {Prsa}, {Quinn}, {Quintana}, {Ragozzine},
  {Ramirez}, {Sanderfer}, {Sasselov}, {Seader}, {Shabram}, {Shporer}, {Smith},
  {Steffen}, {Still}, {Torres}, {Troeltzsch}, {Twicken}, {Uddin}, {Van Cleve},
  {Voss}, {Weiss}, {Welsh}, {Wohler}, \& {Zamudio}}]{Bryson2021}
{Bryson}, S., {Kunimoto}, M., {Kopparapu}, R.~K., {et~al.} 2021, \aj, 161, 36,
  \dodoi{10.3847/1538-3881/abc418}

\bibitem[{{Cowan} {et~al.}(2015){Cowan}, {Greene}, {Angerhausen}, {Batalha},
  {Clampin}, {Col{\'o}n}, {Crossfield}, {Fortney}, {Gaudi}, {Harrington},
  {Iro}, {Lillie}, {Linsky}, {Lopez-Morales}, {Mandell}, \&
  {Stevenson}}]{Cowan2015}
{Cowan}, N.~B., {Greene}, T., {Angerhausen}, D., {et~al.} 2015, \pasp, 127,
  311, \dodoi{10.1086/680855}

\bibitem[{{Deamer} {et~al.}(2022){Deamer}, {Cary}, \& {Damer}}]{Deamer2022}
{Deamer}, D., {Cary}, F., \& {Damer}, B. 2022, Astrobiology, 22, 889,
  \dodoi{10.1089/ast.2021.0173}

\bibitem[{{Dewangan}(2021)}]{Dewangan2021}
{Dewangan}, G.~C. 2021, Journal of Astrophysics and Astronomy, 42, 49,
  \dodoi{10.1007/s12036-021-09691-w}

\bibitem[{{Diamond-Lowe} {et~al.}(2021){Diamond-Lowe}, {Youngblood},
  {Charbonneau}, {King}, {Teal}, {Bastelberger}, {Corrales}, \&
  {Kempton}}]{Diamond-Lowe2021}
{Diamond-Lowe}, H., {Youngblood}, A., {Charbonneau}, D., {et~al.} 2021, \aj,
  162, 10, \dodoi{10.3847/1538-3881/abfa1c}

\bibitem[{{Doherty}(1985)}]{1985MNRAS.217...41D}
{Doherty}, L.~R. 1985, \mnras, 217, 41, \dodoi{10.1093/mnras/217.1.41}

\bibitem[{{Dong} {et~al.}(2017){Dong}, {Lingam}, {Ma}, \& {Cohen}}]{Dong2017}
{Dong}, C., {Lingam}, M., {Ma}, Y., \& {Cohen}, O. 2017, \apjl, 837, L26,
  \dodoi{10.3847/2041-8213/aa6438}

\bibitem[{{Dressing} \& {Charbonneau}(2015)}]{Dressing2015}
{Dressing}, C.~D., \& {Charbonneau}, D. 2015, Astrophysical Journal, 807, 45,
  \dodoi{10.1088/0004-637X/807/1/45}

\bibitem[{{Duvvuri} {et~al.}(2023){Duvvuri}, {Pineda}, {Berta-Thompson},
  {France}, \& {Youngblood}}]{Duvvuri2023}
{Duvvuri}, G.~M., {Pineda}, J.~S., {Berta-Thompson}, Z.~K., {France}, K., \&
  {Youngblood}, A. 2023, \aj, 165, 12, \dodoi{10.3847/1538-3881/ac9b49}

\bibitem[{{Duvvuri} {et~al.}(2021){Duvvuri}, {Sebastian Pineda},
  {Berta-Thompson}, {Brown}, {France}, {Kowalski}, {Redfield}, {Tilipman},
  {Vieytes}, {Wilson}, {Youngblood}, {Froning}, {Linsky}, {Parke Loyd},
  {Mauas}, {Miguel}, {Newton}, {Rugheimer}, \& {Christian
  Schneider}}]{Duvvuri2021}
{Duvvuri}, G.~M., {Sebastian Pineda}, J., {Berta-Thompson}, Z.~K., {et~al.}
  2021, \apj, 913, 40, \dodoi{10.3847/1538-4357/abeaaf}

\bibitem[{France {et~al.}(2013)France, Froning, Linsky, Roberge, Stocke, Tian,
  Bushinsky, D{\'e}sert, Mauas, Vieytes, {et~al.}}]{France2013}
France, K., Froning, C.~S., Linsky, J.~L., {et~al.} 2013, The Astrophysical
  Journal, 763, 149

\bibitem[{{France} {et~al.}(2016){France}, {Parke Loyd}, {Youngblood}, {Brown},
  {Schneider}, {Hawley}, {Froning}, {Linsky}, {Roberge}, {Buccino},
  {Davenport}, {Fontenla}, {Kaltenegger}, {Kowalski}, {Mauas}, {Miguel},
  {Redfield}, {Rugheimer}, {Tian}, {Vieytes}, {Walkowicz}, \&
  {Weisenburger}}]{France2016}
{France}, K., {Parke Loyd}, R.~O., {Youngblood}, A., {et~al.} 2016,
  Astrophysical Journal, 820, 89, \dodoi{10.3847/0004-637X/820/2/89}

\bibitem[{{Gaia Collaboration} {et~al.}(2022){Gaia Collaboration}, {Vallenari},
  {Brown}, {Prusti}, {de Bruijne}, {Arenou}, {Babusiaux}, {Biermann},
  {Creevey}, {Ducourant}, {Evans}, {Eyer}, {Guerra}, {Hutton}, {Jordi},
  {Klioner}, {Lammers}, {Lindegren}, {Luri}, {Mignard}, {Panem}, {Pourbaix},
  {Randich}, {Sartoretti}, {Soubiran}, {Tanga}, {Walton}, {Bailer-Jones},
  {Bastian}, {Drimmel}, {Jansen}, {Katz}, {Lattanzi}, {van Leeuwen}, {Bakker},
  {Cacciari}, {Casta{\~n}eda}, {De Angeli}, {Fabricius}, {Fouesneau},
  {Fr{\'e}mat}, {Galluccio}, {Guerrier}, {Heiter}, {Masana}, {Messineo},
  {Mowlavi}, {Nicolas}, {Nienartowicz}, {Pailler}, {Panuzzo}, {Riclet}, {Roux},
  {Seabroke}, {Sordo{\o}rcit}, {Th{\'e}venin}, {Gracia-Abril}, {Portell},
  {Teyssier}, {Altmann}, {Andrae}, {Audard}, {Bellas-Velidis}, {Benson},
  {Berthier}, {Blomme}, {Burgess}, {Busonero}, {Busso}, {C{\'a}novas}, {Carry},
  {Cellino}, {Cheek}, {Clementini}, {Damerdji}, {Davidson}, {de Teodoro},
  {Nu{\~n}ez Campos}, {Delchambre}, {Dell'Oro}, {Esquej},
  {Fern{\'a}ndez-Hern{\'a}ndez}, {Fraile}, {Garabato}, {Garc{\'\i}a-Lario},
  {Gosset}, {Haigron}, {Halbwachs}, {Hambly}, {Harrison}, {Hern{\'a}ndez},
  {Hestroffer}, {Hodgkin}, {Holl}, {Jan{\ss}en}, {Jevardat de Fombelle},
  {Jordan}, {Krone-Martins}, {Lanzafame}, {L{\"o}ffler}, {Marchal}, {Marrese},
  {Moitinho}, {Muinonen}, {Osborne}, {Pancino}, {Pauwels}, {Recio-Blanco},
  {Reyl{\'e}}, {Riello}, {Rimoldini}, {Roegiers}, {Rybizki}, {Sarro}, {Siopis},
  {Smith}, {Sozzetti}, {Utrilla}, {van Leeuwen}, {Abbas}, {{\'A}brah{\'a}m},
  {Abreu Aramburu}, {Aerts}, {Aguado}, {Ajaj}, {Aldea-Montero}, {Altavilla},
  {{\'A}lvarez}, {Alves}, {Anders}, {Anderson}, {Anglada Varela}, {Antoja},
  {Baines}, {Baker}, {Balaguer-N{\'u}{\~n}ez}, {Balbinot}, {Balog}, {Barache},
  {Barbato}, {Barros}, {Barstow}, {Bartolom{\'e}}, {Bassilana}, {Bauchet},
  {Becciani}, {Bellazzini}, {Berihuete}, {Bernet}, {Bertone}, {Bianchi},
  {Binnenfeld}, {Blanco-Cuaresma}, {Blazere}, {Boch}, {Bombrun}, {Bossini},
  {Bouquillon}, {Bragaglia}, {Bramante}, {Breedt}, {Bressan}, {Brouillet},
  {Brugaletta}, {Bucciarelli}, {Burlacu}, {Butkevich}, {Buzzi}, {Caffau},
  {Cancelliere}, {Cantat-Gaudin}, {Carballo}, {Carlucci}, {Carnerero},
  {Carrasco}, {Casamiquela}, {Castellani}, {Castro-Ginard}, {Chaoul},
  {Charlot}, {Chemin}, {Chiaramida}, {Chiavassa}, {Chornay}, {Comoretto},
  {Contursi}, {Cooper}, {Cornez}, {Cowell}, {Crifo}, {Cropper}, {Crosta},
  {Crowley}, {Dafonte}, {Dapergolas}, {David}, {David}, {de Laverny}, {De
  Luise}, {De March}, {De Ridder}, {de Souza}, {de Torres}, {del Peloso}, {del
  Pozo}, {Delbo}, {Delgado}, {Delisle}, {Demouchy}, {Dharmawardena}, {Di
  Matteo}, {Diakite}, {Diener}, {Distefano}, {Dolding}, {Edvardsson}, {Enke},
  {Fabre}, {Fabrizio}, {Faigler}, {Fedorets}, {Fernique}, {Fienga}, {Figueras},
  {Fournier}, {Fouron}, {Fragkoudi}, {Gai}, {Garcia-Gutierrez},
  {Garcia-Reinaldos}, {Garc{\'\i}a-Torres}, {Garofalo}, {Gavel}, {Gavras},
  {Gerlach}, {Geyer}, {Giacobbe}, {Gilmore}, {Girona}, {Giuffrida}, {Gomel},
  {Gomez}, {Gonz{\'a}lez-N{\'u}{\~n}ez}, {Gonz{\'a}lez-Santamar{\'\i}a},
  {Gonz{\'a}lez-Vidal}, {Granvik}, {Guillout}, {Guiraud},
  {Guti{\'e}rrez-S{\'a}nchez}, {Guy}, {Hatzidimitriou}, {Hauser}, {Haywood},
  {Helmer}, {Helmi}, {Sarmiento}, {Hidalgo}, {Hilger}, {H{\l}adczuk}, {Hobbs},
  {Holland}, {Huckle}, {Jardine}, {Jasniewicz}, {Jean-Antoine Piccolo},
  {Jim{\'e}nez-Arranz}, {Jorissen}, {Juaristi Campillo}, {Julbe}, {Karbevska},
  {Kervella}, {Khanna}, {Kontizas}, {Kordopatis}, {Korn}, {K{\'o}sp{\'a}l},
  {Kostrzewa-Rutkowska}, {Kruszy{\'n}ska}, {Kun}, {Laizeau}, {Lambert},
  {Lanza}, {Lasne}, {Le Campion}, {Lebreton}, {Lebzelter}, {Leccia}, {Leclerc},
  {Lecoeur-Taibi}, {Liao}, {Licata}, {Lindstr{\o}m}, {Lister}, {Livanou},
  {Lobel}, {Lorca}, {Loup}, {Madrero Pardo}, {Magdaleno Romeo}, {Managau},
  {Mann}, {Manteiga}, {Marchant}, {Marconi}, {Marcos}, {Marcos Santos},
  {Mar{\'\i}n Pina}, {Marinoni}, {Marocco}, {Marshall}, {Polo},
  {Mart{\'\i}n-Fleitas}, {Marton}, {Mary}, {Masip}, {Massari},
  {Mastrobuono-Battisti}, {Mazeh}, {McMillan}, {Messina}, {Michalik}, {Millar},
  {Mints}, {Molina}, {Molinaro}, {Moln{\'a}r}, {Monari}, {Mongui{\'o}},
  {Montegriffo}, {Montero}, {Mor}, {Mora}, {Morbidelli}, {Morel}, {Morris},
  {Muraveva}, {Murphy}, {Musella}, {Nagy}, {Noval}, {Oca{\~n}a}, {Ogden},
  {Ordenovic}, {Osinde}, {Pagani}, {Pagano}, {Palaversa}, {Palicio},
  {Pallas-Quintela}, {Panahi}, {Payne-Wardenaar}, {Pe{\~n}alosa Esteller},
  {Penttil{\"a}}, {Pichon}, {Piersimoni}, {Pineau}, {Plachy}, {Plum}, {Poggio},
  {Pr{\v{s}}a}, {Pulone}, {Racero}, {Ragaini}, {Rainer}, {Raiteri}, {Rambaux},
  {Ramos}, {Ramos-Lerate}, {Re Fiorentin}, {Regibo}, {Richards}, {Rios Diaz},
  {Ripepi}, {Riva}, {Rix}, {Rixon}, {Robichon}, {Robin}, {Robin}, {Roelens},
  {Rogues}, {Rohrbasser}, {Romero-G{\'o}mez}, {Rowell}, {Royer}, {Ruz Mieres},
  {Rybicki}, {Sadowski}, {S{\'a}ez N{\'u}{\~n}ez}, {Sagrist{\`a} Sell{\'e}s},
  {Sahlmann}, {Salguero}, {Samaras}, {Sanchez Gimenez}, {Sanna},
  {Santove{\~n}a}, {Sarasso}, {Schultheis}, {Sciacca}, {Segol}, {Segovia},
  {S{\'e}gransan}, {Semeux}, {Shahaf}, {Siddiqui}, {Siebert}, {Siltala},
  {Silvelo}, {Slezak}, {Slezak}, {Smart}, {Snaith}, {Solano}, {Solitro},
  {Souami}, {Souchay}, {Spagna}, {Spina}, {Spoto}, {Steele},
  {Steidelm{\"u}ller}, {Stephenson}, {S{\"u}veges}, {Surdej}, {Szabados},
  {Szegedi-Elek}, {Taris}, {Taylo}, {Teixeira}, {Tolomei}, {Tonello}, {Torra},
  {Torra}, {Torralba Elipe}, {Trabucchi}, {Tsounis}, {Turon}, {Ulla}, {Unger},
  {Vaillant}, {van Dillen}, {van Reeven}, {Vanel}, {Vecchiato}, {Viala},
  {Vicente}, {Voutsinas}, {Weiler}, {Wevers}, {Wyrzykowski}, {Yoldas}, {Yvard},
  {Zhao}, {Zorec}, {Zucker}, \& {Zwitter}}]{GaiaCollaboration2022}
{Gaia Collaboration}, {Vallenari}, A., {Brown}, A.~G.~A., {et~al.} 2022, arXiv
  e-prints, arXiv:2208.00211.
\newblock \doarXiv{2208.00211}

\bibitem[{{Harman} {et~al.}(2018){Harman}, {Felton}, {Hu}, {Domagal-Goldman},
  {Segura}, {Tian}, \& {Kasting}}]{Harman2018}
{Harman}, C.~E., {Felton}, R., {Hu}, R., {et~al.} 2018, \apj, 866, 56,
  \dodoi{10.3847/1538-4357/aadd9b}

\bibitem[{{Hu} {et~al.}(2021){Hu}, {Damiano}, {Scheucher}, {Kite}, {Seager}, \&
  {Rauer}}]{Hu2021waterworlds}
{Hu}, R., {Damiano}, M., {Scheucher}, M., {et~al.} 2021, \apjl, 921, L8,
  \dodoi{10.3847/2041-8213/ac1f92}

\bibitem[{{Hu} {et~al.}(2020){Hu}, {Peterson}, \& {Wolf}}]{Hu2020}
{Hu}, R., {Peterson}, L., \& {Wolf}, E.~T. 2020, \apj, 888, 122,
  \dodoi{10.3847/1538-4357/ab5f07}

\bibitem[{{Hu} {et~al.}(2012){Hu}, {Seager}, \& {Bains}}]{Hu2012}
{Hu}, R., {Seager}, S., \& {Bains}, W. 2012, Astrophysical Journal, 761, 166,
  \dodoi{10.1088/0004-637X/761/2/166}

\bibitem[{{Kasting} {et~al.}(1997){Kasting}, {Whittet}, \&
  {Sheldon}}]{Kasting1997}
{Kasting}, J.~F., {Whittet}, D. C.~B., \& {Sheldon}, W.~R. 1997, Origins of
  Life and Evolution of the Biosphere, 27, 413, \dodoi{10.1023/A:1006596806012}

\bibitem[{{Lehmer} {et~al.}(2020){Lehmer}, {Catling}, \&
  {Krissansen-Totton}}]{Lehmer2020}
{Lehmer}, O.~R., {Catling}, D.~C., \& {Krissansen-Totton}, J. 2020, Nature
  Communications, 11, 6153, \dodoi{10.1038/s41467-020-19896-2}

\bibitem[{{Loyd} \& {France}(2014)}]{Loyd2014}
{Loyd}, R.~O.~P., \& {France}, K. 2014, \apjs, 211, 9,
  \dodoi{10.1088/0067-0049/211/1/9}

\bibitem[{{Loyd} {et~al.}(2018){Loyd}, {Shkolnik}, {Schneider}, {Barman},
  {Meadows}, {Pagano}, \& {Peacock}}]{Loyd2018ApJ...867...70L}
{Loyd}, R.~O.~P., {Shkolnik}, E.~L., {Schneider}, A.~C., {et~al.} 2018, \apj,
  867, 70, \dodoi{10.3847/1538-4357/aae2ae}

\bibitem[{{Loyd} {et~al.}(2016){Loyd}, {France}, {Youngblood}, {Schneider},
  {Brown}, {Hu}, {Linsky}, {Froning}, {Redfield}, {Rugheimer}, \&
  {Tian}}]{Loyd2016}
{Loyd}, R.~O.~P., {France}, K., {Youngblood}, A., {et~al.} 2016, Astrophysical
  Journal, 824, 102, \dodoi{10.3847/0004-637X/824/2/102}

\bibitem[{{Loyd} {et~al.}(2021){Loyd}, {Shkolnik}, {Schneider},
  {Richey-Yowell}, {Jackman}, {Peacock}, {Barman}, {Pagano}, \&
  {Meadows}}]{Loyd2021ApJ...907...91L}
{Loyd}, R.~O.~P., {Shkolnik}, E.~L., {Schneider}, A.~C., {et~al.} 2021, \apj,
  907, 91, \dodoi{10.3847/1538-4357/abd0f0}

\bibitem[{{Luger} \& {Barnes}(2015)}]{Luger2015}
{Luger}, R., \& {Barnes}, R. 2015, Astrobiology, 15, 119,
  \dodoi{10.1089/ast.2014.1231}

\bibitem[{{Lustig-Yaeger} {et~al.}(2019){Lustig-Yaeger}, {Meadows}, \&
  {Lincowski}}]{Lustig-Yaeger2019}
{Lustig-Yaeger}, J., {Meadows}, V.~S., \& {Lincowski}, A.~P. 2019, \aj, 158,
  27, \dodoi{10.3847/1538-3881/ab21e0}

\bibitem[{Lustig-Yaeger {et~al.}(2019)Lustig-Yaeger, Robinson, \&
  Arney}]{Lustig2019coronagraph}
Lustig-Yaeger, J., Robinson, T.~D., \& Arney, G. 2019, Journal of Open Source
  Software, 4, 1387, \dodoi{10.21105/joss.01387}

\bibitem[{{Malo} {et~al.}(2014){Malo}, {Doyon}, {Feiden}, {Albert},
  {Lafreni{\`e}re}, {Artigau}, {Gagn{\'e}}, \& {Riedel}}]{Malo2014}
{Malo}, L., {Doyon}, R., {Feiden}, G.~A., {et~al.} 2014, \apj, 792, 37,
  \dodoi{10.1088/0004-637X/792/1/37}

\bibitem[{{Mamajek} \& {Bell}(2014)}]{Mamajek2014}
{Mamajek}, E.~E., \& {Bell}, C. P.~M. 2014, \mnras, 445, 2169,
  \dodoi{10.1093/mnras/stu1894}

\bibitem[{{Melbourne} {et~al.}(2020){Melbourne}, {Youngblood}, {France},
  {Froning}, {Pineda}, {Shkolnik}, {Wilson}, {Wood}, {Basu}, {Roberge},
  {Schlieder}, {Cauley}, {Loyd}, {Newton}, {Schneider}, {Arulanantham},
  {Berta-Thompson}, {Brown}, {Buccino}, {Kempton}, {Linsky}, {Logsdon},
  {Mauas}, {Pagano}, {Peacock}, {Redfield}, {Rugheimer}, {Schneider}, {Teal},
  {Tian}, {Tilipman}, \& {Vieytes}}]{Melbourne2020}
{Melbourne}, K., {Youngblood}, A., {France}, K., {et~al.} 2020, \aj, 160, 269,
  \dodoi{10.3847/1538-3881/abbf5c}

\bibitem[{{Messina} {et~al.}(2010){Messina}, {Desidera}, {Turatto},
  {Lanzafame}, \& {Guinan}}]{Messina2010}
{Messina}, S., {Desidera}, S., {Turatto}, M., {Lanzafame}, A.~C., \& {Guinan},
  E.~F. 2010, \aap, 520, A15, \dodoi{10.1051/0004-6361/200913644}

\bibitem[{{Morley} {et~al.}(2017){Morley}, {Kreidberg}, {Rustamkulov},
  {Robinson}, \& {Fortney}}]{Morley2017}
{Morley}, C.~V., {Kreidberg}, L., {Rustamkulov}, Z., {Robinson}, T., \&
  {Fortney}, J.~J. 2017, \apj, 850, 121, \dodoi{10.3847/1538-4357/aa927b}

\bibitem[{{National Academies of Sciences Engineering and
  Medicine}(2018)}]{NAP25187}
{National Academies of Sciences Engineering and Medicine}. 2018, Exoplanet
  Science Strategy (Washington, DC: The National Academies Press),
  \dodoi{10.17226/25187}

\bibitem[{{National Academies of Sciences Engineering and
  Medicine}(2021)}]{Decadal2020}
---. 2021, Pathways to Discovery in Astronomy and Astrophysics for the 2020s
  (Washington, DC: The National Academies Press), \dodoi{10.17226/26141}

\bibitem[{{O'Malley-James} \& {Kaltenegger}(2019)}]{O'Malley-James2019}
{O'Malley-James}, J.~T., \& {Kaltenegger}, L. 2019, \mnras, 485, 5598,
  \dodoi{10.1093/mnras/stz724}

\bibitem[{{Peacock} {et~al.}(2019{\natexlab{a}}){Peacock}, {Barman},
  {Shkolnik}, {Hauschildt}, \& {Baron}}]{Peacock2019a}
{Peacock}, S., {Barman}, T., {Shkolnik}, E.~L., {Hauschildt}, P.~H., \&
  {Baron}, E. 2019{\natexlab{a}}, \apj, 871, 235,
  \dodoi{10.3847/1538-4357/aaf891}

\bibitem[{{Peacock} {et~al.}(2019{\natexlab{b}}){Peacock}, {Barman},
  {Shkolnik}, {Hauschildt}, {Baron}, \& {Fuhrmeister}}]{Peacock2019b}
{Peacock}, S., {Barman}, T., {Shkolnik}, E.~L., {et~al.} 2019{\natexlab{b}},
  \apj, 886, 77, \dodoi{10.3847/1538-4357/ab4f6f}

\bibitem[{{Peacock} {et~al.}(2020){Peacock}, {Barman}, {Shkolnik}, {Loyd},
  {Schneider}, {Pagano}, \& {Meadows}}]{Peacock2020}
---. 2020, \apj, 895, 5, \dodoi{10.3847/1538-4357/ab893a}

\bibitem[{{Peacock} {et~al.}(2022){Peacock}, {Barman}, {Schneider}, {Leung},
  {Schwieterman}, {Shkolnik}, \& {Loyd}}]{Peacock2022}
{Peacock}, S., {Barman}, T.~S., {Schneider}, A.~C., {et~al.} 2022, \apj, 933,
  235, \dodoi{10.3847/1538-4357/ac77f2}

\bibitem[{{Pineda} {et~al.}(2021){Pineda}, {Youngblood}, \&
  {France}}]{Pineda2021}
{Pineda}, J.~S., {Youngblood}, A., \& {France}, K. 2021, \apj, 911, 111,
  \dodoi{10.3847/1538-4357/abe8d7}

\bibitem[{{Plavchan} {et~al.}(2020){Plavchan}, {Barclay}, {Gagn{\'e}}, {Gao},
  {Cale}, {Matzko}, {Dragomir}, {Quinn}, {Feliz}, {Stassun}, {Crossfield},
  {Berardo}, {Latham}, {Tieu}, {Anglada-Escud{\'e}}, {Ricker}, {Vanderspek},
  {Seager}, {Winn}, {Jenkins}, {Rinehart}, {Krishnamurthy}, {Dynes}, {Doty},
  {Adams}, {Afanasev}, {Beichman}, {Bottom}, {Bowler}, {Brinkworth}, {Brown},
  {Cancino}, {Ciardi}, {Clampin}, {Clark}, {Collins}, {Davison},
  {Foreman-Mackey}, {Furlan}, {Gaidos}, {Geneser}, {Giddens}, {Gilbert},
  {Hall}, {Hellier}, {Henry}, {Horner}, {Howard}, {Huang}, {Huber}, {Kane},
  {Kenworthy}, {Kielkopf}, {Kipping}, {Klenke}, {Kruse}, {Latouf}, {Lowrance},
  {Mennesson}, {Mengel}, {Mills}, {Morton}, {Narita}, {Newton}, {Nishimoto},
  {Okumura}, {Palle}, {Pepper}, {Quintana}, {Roberge}, {Roccatagliata},
  {Schlieder}, {Tanner}, {Teske}, {Tinney}, {Vanderburg}, {von Braun}, {Walp},
  {Wang}, {Wang}, {Weigand}, {White}, {Wittenmyer}, {Wright}, {Youngblood},
  {Zhang}, \& {Zilberman}}]{Plavchan2020}
{Plavchan}, P., {Barclay}, T., {Gagn{\'e}}, J., {et~al.} 2020, \nat, 582, 497,
  \dodoi{10.1038/s41586-020-2400-z}

\bibitem[{{Postma} \& {Leahy}(2017)}]{postma2017}
{Postma}, J.~E., \& {Leahy}, D. 2017, \pasp, 129, 115002,
  \dodoi{10.1088/1538-3873/aa8800}

\bibitem[{{Ramirez} \& {Kaltenegger}(2014)}]{RamirezKaltenegger2014}
{Ramirez}, R.~M., \& {Kaltenegger}, L. 2014, Astrophysical Journall, 797, L25,
  \dodoi{10.1088/2041-8205/797/2/L25}

\bibitem[{{Ranjan} \& {Sasselov}(2017)}]{Ranjan2017a}
{Ranjan}, S., \& {Sasselov}, D.~D. 2017, Astrobiology, 17, 169,
  \dodoi{10.1089/ast.2016.1519}

\bibitem[{{Ranjan} {et~al.}(2022){Ranjan}, {Seager}, {Zhan}, {Koll}, {Bains},
  {Petkowski}, {Huang}, \& {Lin}}]{Ranjan2022b}
{Ranjan}, S., {Seager}, S., {Zhan}, Z., {et~al.} 2022, \apj, 930, 131,
  \dodoi{10.3847/1538-4357/ac5749}

\bibitem[{Ranjan {et~al.}(2017)Ranjan, Wordsworth, \& Sasselov}]{Ranjan2017c}
Ranjan, S., Wordsworth, R., \& Sasselov, D.~D. 2017, The Astrophysical Journal,
  843, 110.
\newblock \url{http://stacks.iop.org/0004-637X/843/i=2/a=110}

\bibitem[{{Rauer} {et~al.}(2011){Rauer}, {Gebauer}, {Paris}, {Cabrera},
  {Godolt}, {Grenfell}, {Belu}, {Selsis}, {Hedelt}, \& {Schreier}}]{Rauer2011}
{Rauer}, H., {Gebauer}, S., {Paris}, P.~V., {et~al.} 2011, \aap, 529, A8,
  \dodoi{10.1051/0004-6361/201014368}

\bibitem[{Rimmer(2023)}]{Rimmer2023}
Rimmer, P.~B. 2023, Conflicting Models for the Origin of Life, 407

\bibitem[{{Rimmer} {et~al.}(2021{\natexlab{a}}){Rimmer}, {Ranjan}, \&
  {Rugheimer}}]{RimmerRanjan2021}
{Rimmer}, P.~B., {Ranjan}, S., \& {Rugheimer}, S. 2021{\natexlab{a}}, Elements,
  17, 265.
\newblock \doarXiv{2108.08388}

\bibitem[{{Rimmer} {et~al.}(2021{\natexlab{b}}){Rimmer}, {Thompson}, {Xu},
  {Russell}, {Green}, {Ritson}, {Sutherland}, \& {Queloz}}]{Rimmer2021}
{Rimmer}, P.~B., {Thompson}, S.~J., {Xu}, J., {et~al.} 2021{\natexlab{b}},
  Astrobiology, 21, 1099, \dodoi{10.1089/ast.2020.2335}

\bibitem[{{Rimmer} {et~al.}(2018){Rimmer}, {Xu}, {Thompson}, {Gillen},
  {Sutherland}, \& {Queloz}}]{Rimmer2018}
{Rimmer}, P.~B., {Xu}, J., {Thompson}, S.~J., {et~al.} 2018, Science Advances,
  4, eaar3302, \dodoi{10.1126/sciadv.aar3302}

\bibitem[{{Robinson} {et~al.}(2016){Robinson}, {Stapelfeldt}, \&
  {Marley}}]{Robinson2016}
{Robinson}, T.~D., {Stapelfeldt}, K.~R., \& {Marley}, M.~S. 2016, \pasp, 128,
  025003, \dodoi{10.1088/1538-3873/128/960/025003}

\bibitem[{{Rodler} \& {L{\'o}pez-Morales}(2014)}]{Rodler2014}
{Rodler}, F., \& {L{\'o}pez-Morales}, M. 2014, Astrophysical Journal, 781, 54,
  \dodoi{10.1088/0004-637X/781/1/54}

\bibitem[{Rugheimer {et~al.}(2015)Rugheimer, Kaltenegger, Segura, Linsky, \&
  Mohanty}]{Rugheimer2015mdwarf}
Rugheimer, S., Kaltenegger, L., Segura, A., Linsky, J., \& Mohanty, S. 2015,
  The Astrophysical Journal, 809, 57

\bibitem[{{Rugheimer} {et~al.}(2015){Rugheimer}, {Segura}, {Kaltenegger}, \&
  {Sasselov}}]{Rugheimer2015}
{Rugheimer}, S., {Segura}, A., {Kaltenegger}, L., \& {Sasselov}, D. 2015,
  Astrophysical Journal, 806, 137, \dodoi{10.1088/0004-637X/806/1/137}

\bibitem[{{Scalo} {et~al.}(2007){Scalo}, {Kaltenegger}, {Segura}, {Fridlund},
  {Ribas}, {Kulikov}, {Grenfell}, {Rauer}, {Odert}, {Leitzinger}, {Selsis},
  {Khodachenko}, {Eiroa}, {Kasting}, \& {Lammer}}]{Scalo2007}
{Scalo}, J., {Kaltenegger}, L., {Segura}, A.~G., {et~al.} 2007, Astrobiology,
  7, 85, \dodoi{10.1089/ast.2006.0125}

\bibitem[{{Schwieterman} {et~al.}(2019){Schwieterman}, {Reinhard}, {Olson},
  {Ozaki}, {Harman}, {Hong}, \& {Lyons}}]{Schwieterman2019}
{Schwieterman}, E.~W., {Reinhard}, C.~T., {Olson}, S.~L., {et~al.} 2019, \apj,
  874, 9, \dodoi{10.3847/1538-4357/ab05e1}

\bibitem[{{Schwieterman} {et~al.}(2018){Schwieterman}, {Kiang}, {Parenteau},
  {Harman}, {DasSarma}, {Fisher}, {Arney}, {Hartnett}, {Reinhard}, {Olson},
  {Meadows}, {Cockell}, {Walker}, {Grenfell}, {Hegde}, {Rugheimer}, {Hu}, \&
  {Lyons}}]{Schwieterman2018}
{Schwieterman}, E.~W., {Kiang}, N.~Y., {Parenteau}, M.~N., {et~al.} 2018,
  Astrobiology, 18, 663, \dodoi{10.1089/ast.2017.1729}

\bibitem[{{Schwieterman} {et~al.}(2022){Schwieterman}, {Olson},
  {Pidhorodetska}, {Reinhard}, {Ganti}, {Fauchez}, {Bastelberger}, {Crouse},
  {Ridgwell}, \& {Lyons}}]{Schwieterman2022}
{Schwieterman}, E.~W., {Olson}, S.~L., {Pidhorodetska}, D., {et~al.} 2022,
  \apj, 937, 109, \dodoi{10.3847/1538-4357/ac8cfb}

\bibitem[{{Seager} {et~al.}(2013){Seager}, {Bains}, \& {Hu}}]{Seager2013b}
{Seager}, S., {Bains}, W., \& {Hu}, R. 2013, \apj, 777, 95,
  \dodoi{10.1088/0004-637X/777/2/95}

\bibitem[{{Segura} {et~al.}(2005){Segura}, {Kasting}, {Meadows}, {Cohen},
  {Scalo}, {Crisp}, {Butler}, \& {Tinetti}}]{Segura2005}
{Segura}, A., {Kasting}, J.~F., {Meadows}, V., {et~al.} 2005, Astrobiology, 5,
  706, \dodoi{10.1089/ast.2005.5.706}

\bibitem[{{Shields} {et~al.}(2016){Shields}, {Ballard}, \&
  {Johnson}}]{Shields2016rev}
{Shields}, A.~L., {Ballard}, S., \& {Johnson}, J.~A. 2016, ArXiv e-prints.
\newblock \doarXiv{1610.05765}

\bibitem[{{Shkolnik} \& {Barman}(2014)}]{Shkolnik2014}
{Shkolnik}, E.~L., \& {Barman}, T.~S. 2014, \aj, 148, 64,
  \dodoi{10.1088/0004-6256/148/4/64}

\bibitem[{{Subramaniam}(2022)}]{Subramaniam2022}
{Subramaniam}, A. 2022, Journal of Astrophysics and Astronomy, 43, 80,
  \dodoi{10.1007/s12036-022-09870-3}

\bibitem[{{Tandon} {et~al.}(2017){Tandon}, {Subramaniam}, {Girish}, {Postma},
  {Sankarasubramanian}, {Sriram}, {Stalin}, {Mondal}, {Sahu}, {Joseph},
  {Hutchings}, {Ghosh}, {Barve}, {George}, {Kamath}, {Kathiravan}, {Kumar},
  {Lancelot}, {Leahy}, {Mahesh}, {Mohan}, {Nagabhushana}, {Pati}, {Kameswara
  Rao}, {Sreedhar}, \& {Sreekumar}}]{Tandon2017}
{Tandon}, S.~N., {Subramaniam}, A., {Girish}, V., {et~al.} 2017, \aj, 154, 128,
  \dodoi{10.3847/1538-3881/aa8451}

\bibitem[{{Tandon} {et~al.}(2020){Tandon}, {Postma}, {Joseph}, {Devaraj},
  {Subramaniam}, {Barve}, {George}, {Ghosh}, {Girish}, {Hutchings}, {Kamath},
  {Kathiravan}, {Kumar}, {Lancelot}, {Leahy}, {Mahesh}, {Mohan},
  {Nagabhushana}, {Pati}, {Rao}, {Sankarasubramanian}, {Sriram}, \&
  {Stalin}}]{tandon2020}
{Tandon}, S.~N., {Postma}, J., {Joseph}, P., {et~al.} 2020, \aj, 159, 158,
  \dodoi{10.3847/1538-3881/ab72a3}

\bibitem[{{Teal} {et~al.}(2022){Teal}, {Kempton}, {Bastelberger}, {Youngblood},
  \& {Arney}}]{Teal2022}
{Teal}, D.~J., {Kempton}, E. M.~R., {Bastelberger}, S., {Youngblood}, A., \&
  {Arney}, G. 2022, arXiv e-prints, arXiv:2201.08805.
\newblock \doarXiv{2201.08805}

\bibitem[{{Thompson} {et~al.}(2022){Thompson}, {Krissansen-Totton}, {Wogan},
  {Telus}, \& {Fortney}}]{Thompson2022}
{Thompson}, M.~A., {Krissansen-Totton}, J., {Wogan}, N., {Telus}, M., \&
  {Fortney}, J.~J. 2022, Proceedings of the National Academy of Science, 119,
  e2117933119, \dodoi{10.1073/pnas.2117933119}

\bibitem[{{Tilley} {et~al.}(2019){Tilley}, {Segura}, {Meadows}, {Hawley}, \&
  {Davenport}}]{Tilley2019}
{Tilley}, M.~A., {Segura}, A., {Meadows}, V., {Hawley}, S., \& {Davenport}, J.
  2019, Astrobiology, 19, 64, \dodoi{10.1089/ast.2017.1794}

\bibitem[{{Torres} {et~al.}(2006){Torres}, {Quast}, {da Silva}, {de La Reza},
  {Melo}, \& {Sterzik}}]{Torres2006}
{Torres}, C.~A.~O., {Quast}, G.~R., {da Silva}, L., {et~al.} 2006, \aap, 460,
  695, \dodoi{10.1051/0004-6361:20065602}

\bibitem[{{Tremblay} {et~al.}(2020){Tremblay}, {Line}, {Stevenson}, {Kataria},
  {Zellem}, {Fortney}, \& {Morley}}]{Tremblay2020AJ....159..117T}
{Tremblay}, L., {Line}, M.~R., {Stevenson}, K., {et~al.} 2020, \aj, 159, 117,
  \dodoi{10.3847/1538-3881/ab64dd}

\bibitem[{{Viswanath} {et~al.}(2020){Viswanath}, {Narang}, {Manoj}, {Mathew},
  \& {Kartha}}]{Viswanath2020}
{Viswanath}, G., {Narang}, M., {Manoj}, P., {Mathew}, B., \& {Kartha}, S.~S.
  2020, \aj, 159, 194, \dodoi{10.3847/1538-3881/ab7d3b}

\bibitem[{{Walkowicz} {et~al.}(2008){Walkowicz}, {Johns-Krull}, \&
  {Hawley}}]{Walkowicz2008}
{Walkowicz}, L.~M., {Johns-Krull}, C.~M., \& {Hawley}, S.~L. 2008, \apj, 677,
  593, \dodoi{10.1086/526421}

\bibitem[{{Wogan} \& {Catling}(2020)}]{Wogan2020}
{Wogan}, N.~F., \& {Catling}, D.~C. 2020, \apj, 892, 127,
  \dodoi{10.3847/1538-4357/ab7b81}

\bibitem[{{Wordsworth} {et~al.}(2018){Wordsworth}, {Schaefer}, \&
  {Fischer}}]{Wordsworth2018}
{Wordsworth}, R.~D., {Schaefer}, L.~K., \& {Fischer}, R.~A. 2018, \aj, 155,
  195, \dodoi{10.3847/1538-3881/aab608}

\bibitem[{Xu {et~al.}(2018)Xu, Ritson, Ranjan, Todd, Sasselov, \&
  Sutherland}]{Xu2018}
Xu, J., Ritson, D.~J., Ranjan, S., {et~al.} 2018, Chemical Communications

\bibitem[{{Xu} {et~al.}(2020){Xu}, {Chmela}, {Green}, {Russell}, {Janicki},
  {G{\'o}ra}, {Szabla}, {Bond}, \& {Sutherland}}]{Xu2020}
{Xu}, J., {Chmela}, V., {Green}, N.~J., {et~al.} 2020, \nat, 582, 60,
  \dodoi{10.1038/s41586-020-2330-9}

\bibitem[{{Youngblood} {et~al.}(2021){Youngblood}, {Pineda}, \&
  {France}}]{Youngblood2021}
{Youngblood}, A., {Pineda}, J.~S., \& {France}, K. 2021, \apj, 911, 112,
  \dodoi{10.3847/1538-4357/abe8d8}

\bibitem[{{Youngblood} {et~al.}(2016){Youngblood}, {France}, {Loyd}, {Linsky},
  {Redfield}, {Schneider}, {Wood}, {Brown}, {Froning}, {Miguel}, {Rugheimer},
  \& {Walkowicz}}]{Youngblood2016}
{Youngblood}, A., {France}, K., {Loyd}, R.~O.~P., {et~al.} 2016, \apj, 824,
  101, \dodoi{10.3847/0004-637X/824/2/101}

\bibitem[{{Youngblood} {et~al.}(2017){Youngblood}, {France}, {Loyd}, {Brown},
  {Mason}, {Schneider}, {Tilley}, {Berta-Thompson}, {Buccino}, {Froning},
  {Hawley}, {Linsky}, {Mauas}, {Redfield}, {Kowalski}, {Miguel}, {Newton},
  {Rugheimer}, {Segura}, {Roberge}, \& {Vieytes}}]{Youngblood2017}
---. 2017, \apj, 843, 31, \dodoi{10.3847/1538-4357/aa76dd}

\end{thebibliography}
\bibliographystyle{aasjournal}



\end{document}